\documentclass[11pt,a4paper]{article}
\usepackage{cite}
\usepackage{graphicx}
\usepackage{amssymb}
\usepackage{amsmath}
\usepackage{amsfonts}
\usepackage{dsfont}
\usepackage{mathtools}
\usepackage{array}
\usepackage{rotating}
\usepackage{bbold,amsfonts}

\usepackage[utf8]{inputenc}
\usepackage{bm}
\usepackage{xcolor}
\usepackage{float}
\usepackage{braket}

\usepackage[height=8.8in,width=6.45in]{geometry}
\usepackage[font=small,labelfont=bf]{caption}
\usepackage[hidelinks]{hyperref}
\usepackage{booktabs,float,slashed}
\usepackage{standalone}
\usepackage{caption}
\usepackage{subcaption}
\numberwithin{equation}{section}
\newcommand{\vev}[1]{{\left\langle #1 \right\rangle}}

\newcommand{\beq}{\begin{equation}}
\newcommand{\eeq}{\end{equation}}

\newcommand{\overbar}[1]{\mkern 1.5mu\overline{\mkern-1.5mu#1\mkern-1.5mu}\mkern 1.5mu}

\DeclareMathOperator{\Tr}{Tr}
\DeclareMathOperator{\tr}{tr}

\makeatletter
\newcommand*{\letterdef@}{}
\newcommand*{\letterdef}[3]{%
	\def\letterdef@##1{\expandafter\newcommand\csname #1\endcsname{#2{##1}}}%
	\@tfor\@tempa :=#3\do{\expandafter\letterdef@\expandafter{\@tempa}}}
\makeatother
\letterdef{c#1} {\mathcal}{ABCDEFGHIJKLMNOPQRSTUVWXYZ} 
\letterdef{rm#1}{\mathrm} {dDeimM} 

\newcommand{\Xx}{\mathsf{X}}

\newdimen\tableauside\tableauside=1.0ex
\newdimen\tableaurule\tableaurule=0.4pt
\newdimen\tableaustep
\def\phantomhrule#1{\hbox{\vbox to0pt{\hrule height\tableaurule
			width#1\vss}}}
\def\phantomvrule#1{\vbox{\hbox to0pt{\vrule width\tableaurule
			height#1\hss}}}
\def\sqr{\vbox{%
		\phantomhrule\tableaustep
		\hbox{\phantomvrule\tableaustep\kern\tableaustep\phantomvrule\tableaustep}%
		\hbox{\vbox{\phantomhrule\tableauside}\kern-\tableaurule}}}
\def\squares#1{\hbox{\count0=#1\noindent\loop\sqr
		\advance\count0 by-1 \ifnum\count0>0\repeat}}
\def\tableau#1{\vcenter{\offinterlineskip
		\tableaustep=\tableauside\advance\tableaustep by-\tableaurule
		\kern\normallineskip\hbox
		{\kern\normallineskip\vbox
			{\gettableau#1 0 }%
			\kern\normallineskip\kern\tableaurule}%
		\kern\normallineskip\kern\tableaurule}}
\def\gettableau#1 {\ifnum#1=0\let\next=\null\else
	\squares{#1}\let\next=\gettableau\fi\next}
\allowdisplaybreaks

\begin{document}
\begin{titlepage}
\vspace*{10mm}
\begin{center}
{\LARGE \bf 
	 	Three-point functions in a $\cN=2$ superconformal gauge theory and their strong-coupling limit
}

\vspace*{15mm}

{\Large M. Bill\`o${}^{\,a,c}$, M. Frau${}^{\,a,c}$, A. Lerda${}^{\,b,c}$, A. Pini${}^{\,c}$, P. Vallarino${}^{\,a,c}$}

\vspace*{8mm}
	
${}^a$ Universit\`a di Torino, Dipartimento di Fisica,\\
			Via P. Giuria 1, I-10125 Torino, Italy
			\vskip 0.3cm

${}^b$  Universit\`a del Piemonte Orientale,\\
			Dipartimento di Scienze e Innovazione Tecnologica\\
			Viale T. Michel 11, I-15121 Alessandria, Italy
			\vskip 0.3cm
			
${}^c$   I.N.F.N. - sezione di Torino,\\
			Via P. Giuria 1, I-10125 Torino, Italy

\vskip 0.8cm
	{\small
		E-mail:
		\texttt{billo,frau,lerda,apini,vallarin@to.infn.it}
	}
\vspace*{0.8cm}
\end{center}

\begin{abstract}
	We study the 3-point functions of single-trace scalar operators in a four-dimensional $\cN=2$ SYM theory with gauge group $\mathrm{SU}(N)$ and matter in the symmetric plus anti-symmetric representation, which has a vanishing $\beta$-function. By mapping this computation to the matrix model arising from localization on a 4-sphere we are able to resum the perturbative expansion in the large-$N$ 't Hooft limit and derive the behavior of the correlators at strong coupling. Finally, by combining our results on the 3-point functions with those on the 2-point functions that have been recently found, we obtain the normalized 3-point coefficients of this conformal field theory at strong coupling and find that they depend in a simple way on the conformal dimensions of the single-trace operators.

\end{abstract}
\end{titlepage}
\setcounter{tocdepth}{2}
\tableofcontents
\vspace{1cm}
\section{Introduction}
\label{sec:intro}

Four-dimensional gauge theories with extended supersymmetry are a typical 
playground where to
find and test techniques that can shed light on the strong-coupling regime. A lot of 
progress in this direction has been made over the years in the maximally supersymmetric 
theory, {\it{i.e.}} $\mathcal{N}=4$ Super Yang-Mills (SYM), especially in the planar limit 
of a large number of colors. In this case a variety of methods, like for instance 
localization, integrability, holography and others, have been successfully used to obtain 
information on the strong-coupling phase of the theory.

When the supersymmetry is not maximal, things are more complicated. In the last few 
years, however, significant developments have been realized in the context of 
$\mathcal{N}=2$ gauge theories with the use of localization techniques (for a review see 
for example \cite{Pestun:2016jze}). Indeed, as originally shown in \cite{Pestun:2007rz}, a 
generic $\mathcal{N}=2$ SYM theory in flat space can be mapped 
to a matrix model defined on a 4-sphere and the functional path-integral can be reduced
to a finite dimensional integration over the elements of a matrix. 
Using this approach, many interesting results have
been obtained in particular when the $\mathcal{N}=2$ theory is superconformal\,%
\footnote{$\mathcal{N}=2$ superconformal gauge theories were originally investigated in \cite{Howe:1983wj}.}, like
for example the Wilson loop vacuum expectation value 
\cite{Andree:2010na,Rey:2010ry,Passerini:2011fe,Russo:2012ay,Russo:2017ngf,Beccaria:2021ksw,Beccaria:2021vuc,Beccaria:2021ism}, the chiral/anti-chiral correlators \cite{Baggio:2014sna,Baggio:2015vxa,Gerchkovitz:2016gxx,Baggio:2016skg,Rodriguez-Gomez:2016ijh,Rodriguez-Gomez:2016cem,Pini:2017ouj,Billo:2017glv,Bourget:2018obm,Beccaria:2018xxl,Billo:2019fbi,Beccaria:2020azj,Beccaria:2020hgy,Galvagno:2020cgq,Beccaria:2021hvt,Fiol:2021icm,Billo:2021rdb}, the correlators of chiral 
operators and Wilson loops \cite{Semenoff:2001xp,Billo:2018oog,Beccaria:2018owt,Beccaria:2020ykg,Galvagno:2021bbj}, the free energy \cite{Fiol:2020bhf,Fiol:2020ojn,Fiol:2021jsc} and the Bremsstrahlung function \cite{Fiol:2015spa,Fiol:2015mrp,Bianchi:2018zpb,Bianchi:2019dlw,Galvagno:2021qyq}. In the weak-coupling regime it is possible to check at the first perturbative orders that the results obtained with the matrix model agree with those obtained with standard Feynman diagrams (see for example \cite{Andree:2010na,Billo:2017glv,Gomez:2018usu,Billo:2019job,Billo:2019fbi}). However, while the diagrammatic methods soon become unpractical, the matrix model approach allows one to obtain explicit results with little computational effort
even at high orders in perturbation theory. In this way one can efficiently generate long series expansions that are very useful for numerical simulations.

These calculations become particularly simple in a special $\mathcal{N}=2$ SYM theory whose matter hypermultiplets transform in the symmetric plus anti-symmetric representation. This theory, which has a vanishing $\beta$-function, has been dubbed ``$\mathbf{E}$ theory'' in \cite{Billo:2019fbi,Beccaria:2020hgy} and represents the $\mathcal{N}=2$ gauge theory which is closest to the $\mathcal{N}=4$ SYM, in the sense that it shares with it many properties even though it has only half of the maximal supersymmetry. The main reason behind this fact is that the hypermultiplets of the $\mathbf{E}$ theory are altogether in a representation
(symmetric plus anti-symmetric) which is not so different from the adjoint representation to which the hypermultiplets of the $\mathcal{N}=4$ SYM belong.
This similarity becomes more evident in the planar limit where several observables, 
like for instance the free energy and the vacuum expectation value of the circular Wilson loop,
coincide in the two theories and the differences show up only in the non-planar sector. However, the $\mathcal{N}=4$ SYM and the $\mathbf{E}$ theory are not planar equivalent since there are other observables, like for instance the correlators of gauge invariant operators of odd conformal dimensions, which remain different even in the planar approximation and are therefore very interesting to study.
These features have a nice interpretation in the dual holographic description. Indeed, 
the $\mathbf{E}$ theory is dual to Type II B string theory in $\mathrm{AdS}_5\times S^5/\mathbb{Z}_2$ (see for example \cite{Ennes:2000fu}) which is realized as
a suitable $\mathbb{Z}_2$ orbifold/orientifold projection of
$\mathrm{AdS}_5\times S^5$ that is the well-known holographic dual of the
$\mathcal{N}=4$ SYM \cite{Maldacena:1997re}. Therefore, all observables of the 
$\mathbf{E}$ theory which in the holographic dictionary correspond to strings 
excitations of the untwisted sector and are thus
insensitive to the orbifold/orientifold, must coincide at strong coupling
with those of the maximally supersymmetric theory in the planar limit.
On the contrary, the observables of the $\mathbf{E}$ theory which correspond to string configurations of the twisted sector crucially feel the presence of the
orbifold/orientifold and at strong coupling deviate from those of the $\mathcal{N}=4$ SYM even in the planar limit.

In this paper we continue the study of the $\mathbf{E}$ theory with group SU($N$) which was initiated in \cite{Beccaria:2020hgy,Beccaria:2021hvt}, 
with the aim of making a further step towards a complete understanding of its strong coupling regime. To do so we exploit the power of the matrix model which, as mentioned above, allows us to obtain explicit expressions for many observables at a small computational cost. While the matrix model associated to the
$\mathcal{N}=4$ SYM is free, the one corresponding to the $\mathbf{E}$ theory
is interacting but with an interaction action that is remarkably simple despite the fact that it contains an infinite number of terms. Further simplifications occur in the large-$N$ limit where one is able to resum the (long) perturbative expansions produced by
the matrix model and infer from them formal expressions that are valid for \emph{all} values of the 't Hooft coupling $\lambda$. In particular one can show \cite{Beccaria:2020hgy,Beccaria:2021hvt} that in the planar limit the partition function $\mathcal{Z}$
of the matrix model acquires the form
\begin{equation}
\mathcal{Z}=\det{}^{-\frac{1}{2}}\big(\mathbb{1}-\Xx\big)
\end{equation}
where $\Xx$ is an infinite $\lambda$-dependent matrix whose elements are
known in terms of an integral convolution of Bessel functions with arguments proportional to $\sqrt{\lambda}$. Expanding these Bessel functions in power series
for small values of $\lambda$ one recovers the perturbative results at weak coupling, while if one uses the asymptotic limit of the Bessel functions for large values of $\lambda$ one can obtain the strong-coupling behavior. Using this method in \cite{Beccaria:2021hvt} the 2-point functions of the single-trace chiral/anti-chiral operators of the
$\mathbf{E}$ theory have been studied in detail both at week and at strong coupling in the planar limit. In particular, the 2-point correlators of operators with even conformal dimensions, which in the string construction belong to the untwisted sector, do not receive $\lambda$-dependent corrections in the planar limit and coincide with the corresponding ones of the $\mathcal{N}=4$ SYM. On the contrary, the 2-point functions of operators with odd conformal dimension, which correspond to string configurations of
the twisted sector, deviate from those of the $\mathcal{N}=4$ SYM even in the planar approximation and at strong coupling are proportional to $1/\lambda$. Similar results have been obtained in \cite{Billo:2021rdb} for quiver theories, from which the $\mathbf{E}$ theory descends with a suitable projection.

Here we generalize this analysis to the 3-point functions of single-trace scalar operators\,%
\footnote{While the matrix-model techniques are completely general and can be applied also to multi-trace operators, we focus on the single-trace operators since in the large-$N$ limit they form a closed set of observables, in the sense that their planar 3-point functions do not involve mixings with multi-trace operators \cite{Baggio:2016skg}.}. It is worth recalling that so far the 3-point functions in $\mathcal{N}=2$ SYM theories have not been considered very much in the matrix-model literature. Indeed, one can find only some results for the 3-point correlators of operators with even dimension in the SU(2) $\mathcal{N}=2$ superconformal QCD in \cite{Baggio:2014sna}, or some perturbative results for such correlators at large $N$ in \cite{Baggio:2016skg} and more recently in \cite{Fiol:2021icm}, where also a resummation of all terms linear in the Riemann $\zeta$-values has been proposed. In this paper we fill this gap and study in detail the 3-point functions of single-trace scalar operators in the $\mathbf{E}$ theory both at weak and at strong coupling. Exploiting the simplicity of the
matrix model of this theory we are able to analyze the 3-point functions in full generality and find that only the correlators involving two operators with odd conformal dimension deviate from the $\mathcal{N}=4$ SYM expressions and become proportional 
to $1/\lambda$ at strong coupling. To our knowledge this is the first explicit 
result on the 3-point functions in a $\mathcal{N}=2$ SYM theory in which the perturbative expansion has been fully resummed and extrapolated at strong coupling. Combining these findings with those on the 2-point functions, we also compute the normalized 3-point coefficients at strong coupling and find that they depend on the conformal dimensions of the operators in a remarkably simple way that is similar, but of course not identical, to that of the $\mathcal{N}=4$ SYM in the planar limit. In particular, 
we find that the normalized coefficient in the correlation function of three untwisted operators at strong coupling in the planar limit is 
\begin{equation}
C_{U_1U_2U_3}=\frac{1}{N}\,\sqrt{\phantom{\big|}\!d_{U_1}\,d_{U_2}\,d_{U_3} }
\label{CUUU0}
\end{equation}
where the $d$'s are the conformal dimensions of the operators, and that the normalized coefficient in the 3-point function
of one untwisted and two twisted operators is 
\begin{equation}
C_{U_1T_2T_3}=\frac{1}{N}\,\sqrt{\phantom{\big|}\!d_{U_1}\,(d_{T_2}-1)\,
(d_{T_3}-1)}~.
\label{CUTT0}
\end{equation}
These formulas are exact at large $N$ and receive corrections at order 
$O\big(1/N^3\big)$. With these normalized coefficients one can in principle obtain other
correlation functions at strong coupling 
using the standard conformal field theory methods.
We thus believe that our findings are a significant step towards a more complete understanding of the strong-coupling regime in the case of conformal theories with $\mathcal{N}=2$ supersymmetry.

The paper is organized as follows: in Section~\ref{secn:gauge} we review the main features of the extremal correlators in a generic $\mathcal{N}=2$ superconformal gauge theory with group SU($N$) and of their computation in the matrix model using localization. In Section~\ref{secn:Etheory} we focus on the $\mathbf{E}$ theory and exhibit the first perturbative terms of the 2- and 3-point correlators of single-trace operators at large $N$, providing also a simple interpretation in terms of Feynman diagrams. In Section~\ref{secn:example} we present a complete analysis of the simplest  3-point function, showing in particular how to obtain its strong-coupling limit. We also perform some numerical checks and compare our results with a Monte Carlo simulation. In Section~\ref{secn:Estrong} we discuss the most general 3-point function of single-trace operators of the $\mathbf{E}$ theory at large $N$ and at strong coupling. Finally in
Section~\ref{secn:concl}, after deriving the normalized 3-point coefficients and their
dependence on the conformal dimensions of the operators, we draw our conclusions.
Several technical details that are useful to check and reproduce our calculations are collected in the appendices, which contain also various formulas and results that may be relevant for possible extensions of our analysis beyond the leading planar approximation.

\section{Extremal correlators in \texorpdfstring{$\mathcal{N}=2$}{} superconformal gauge theories}
\label{secn:gauge}
We consider a generic $\mathcal{N}=2$ superconformal Yang-Mills theory 
in $\mathbb{R}^4$ with gauge group SU($N$) and denote by $\varphi(x)$ the complex field in the adjoint vector multiplet. A set of interesting gauge-invariant local operators of this theory is that of
the multi-traces of the powers of $\varphi$. Given a collection of integers
\begin{equation}
\mathbf{n}=\{n_1,n_2,\ldots,n_\ell\}~,
\label{n}
\end{equation}
we define
\begin{equation}
\mathcal{O}_{\mathbf{n}}(x)=\tr \varphi(x)^{n_1}\,\tr \varphi(x)^{n_2}\ldots\tr
\varphi(x)^{n_\ell}
\label{Onx}
\end{equation}
where the traces are in the fundamental representation.
$\mathcal{O}_{\mathbf{n}}(x)$ is a chiral primary operator with conformal dimension
\begin{equation}
|\mathbf{n}|=n_1+n_2+\cdots +n_\ell~,
\label{dim}
\end{equation}
and is annihilated by half of the supercharges of the $\mathcal{N}=2$ algebra.
The anti-chiral operator $\overbar{\mathcal{O}}_{\mathbf{n}}(x)$
is defined in a similar way with $\varphi(x)$ replaced by its complex conjugate $\overbar{\varphi}(x)$, and is annihilated by the other half of the supercharges.

The operators (\ref{Onx}) satisfy the freely generated chiral ring relation
\begin{equation}
\mathcal{O}_{\mathbf{n}}(x)\,\mathcal{O}_{\mathbf{m}}(0)=
\mathcal{O}_{\mathbf{n},\mathbf{m}}(0)+\ldots~,
\label{chiralring}
\end{equation}
where the notation $\mathbf{n},\mathbf{m}$ simply denotes the union of $\mathbf{n}$ and $\mathbf{m}$, and the ellipses stand for terms that are exact with respect to the supercharges annihilating the chiral operators.

In the following we will study the so-called ``extremal'' correlators given by
\begin{equation}
\big\langle \,\mathcal{O}_{\mathbf{n_1}}(x_1)\ldots
\,\mathcal{O}_{\mathbf{n_k}}(x_k)\,\overbar{\mathcal{O}}_{\mathbf{m}}(y)\,\big\rangle
=\frac{\mathcal{G}_{\mathbf{n_1},\ldots,\,\mathbf{n_k};\mathbf{m}}\phantom{\big|}}{
\big(4\pi^2(x_1-y)^2\big)^{|\mathbf{n_1}|}\ldots\big(4\pi^2(x_k-y)^2\big)^{|\mathbf{n_k}|}}
\label{extremal}
\end{equation}
where the space-dependent terms in the denominator arise from the free scalar propagator in $\mathbb{R}^4$ and the coefficient $\mathcal{G}_{\mathbf{n_1},\ldots,\,\mathbf{n_k};\mathbf{m}}$ in the numerator is a non-trivial function of the Yang-Mills coupling $g$ and of $N$ which satisfies the selection rule
\begin{equation}
\mathcal{G}_{\mathbf{n_1},\ldots,\,\mathbf{n_k};\mathbf{m}}~\propto~
\delta_{|\mathbf{n_1}|+\cdots+|\mathbf{n_k}|,|\mathbf{m}|}
\label{selection}
\end{equation}
imposed by the U(1)$_R$ symmetry.
By repeatedly using the chiral ring relation (\ref{chiralring}), it is easy to realize that this same coefficient appears in the 2-point function
\begin{equation}
\big\langle \,\mathcal{O}_{\mathbf{n_1},\ldots,\mathbf{n_k}}(x)\,\overbar{\mathcal{O}}_{\mathbf{m}}(y)\,\big\rangle
=\frac{\mathcal{G}_{\mathbf{n_1},\ldots,\,\mathbf{n_k};\mathbf{m}}\phantom{\big|}}{
\big(4\pi^2(x-y)^2\big)^{|\mathbf{n_1}|+\cdots+|\mathbf{n_k}|}}~.
\label{2pointgen}
\end{equation}
Thus, the computation of the extremal correlators (\ref{extremal}) is effectively reduced to the computation of the 2-point functions of generic multi-trace operators.

Particular cases of the general formula (\ref{2pointgen}), on which we will focus in the following
sections, are the 2-point functions of single-trace operators 
\begin{equation}
\big\langle \,\mathcal{O}_{n}(x)\,\overbar{\mathcal{O}}_{m}(y)\,\big\rangle
=\frac{\mathcal{G}_{n;m}\phantom{\big|}}{
\big(4\pi^2(x-y)^2\big)^{n}}
\label{2pointsingle}
\end{equation}
where
\begin{equation}
\mathcal{G}_{n;m}=G_n\,\delta_{n,m}~,
\label{Gnis}
\end{equation}
and the 3-point functions of single-trace operators
\begin{equation}
\big\langle \,\mathcal{O}_{n_1}(x_1)\,
\mathcal{O}_{n_2}(x_2)\,\overbar{\mathcal{O}}_{m}(y)\,\big\rangle
=\frac{\mathcal{G}_{n_1,n_2;m}\phantom{\big|}}{
\big(4\pi^2(x_1-y)^2\big)^{n_1}\,\big(4\pi^2(x_2-y)^2\big)^{n_2}}
\label{3pointsingle}
\end{equation}
where
\begin{equation}
\mathcal{G}_{n_1,n_2;m}=G_{n_1,n_2}\,\delta_{n_1+n_2,m}\quad\mbox{with}\quad G_{n_1,n_2}=G_{n_2,n_1}~.
\label{Gn1n2is}
\end{equation}

The coefficients $\mathcal{G}_{\mathbf{n_1},\ldots,\,\mathbf{n_k};\mathbf{m}}$ can in principle be computed in perturbation theory using Feynman diagrams, but in this way only very few terms can be found in an explicit form due to the intrinsic difficulty of the evaluation of the loop integrals. As originally pointed out in \cite{Baggio:2014sna,Baggio:2015vxa,Gerchkovitz:2016gxx,Baggio:2016skg}, a much more efficient way to obtain these coefficients, even at high orders, is by using the localization techniques and matrix models.

\subsection{Matrix model approach}
\label{subsecn:matrix}
As extensively discussed in the literature (for a review see for example \cite{Pestun:2016jze}), by exploiting localization one can replace a $\mathcal{N}=2$ SYM theory on $\mathbb{R}^4$ with an interacting matrix model on a 4-sphere $S^4$ \cite{Pestun:2007rz} and reduce the calculation of the correlation functions to finite dimensional matrix integrals.

Denoting by $a$ a ($N\times N$) traceless hermitian matrix, such that
\begin{equation}
a= a^b\,T_b 
\label{amatrix}
\end{equation}
where $T_b$ ($b=1,\ldots,N^2-1$) are the SU($N$) generators in the fundamental representation\,%
\footnote{Here and in the following we fix the normalization of the generators $T_b$ in such a way that $\tr T_b\,T_c=\frac{1}{2}\delta_{bc}$.}, the partition function of the matrix model is given by
\begin{equation}
\mathcal{Z}=\int da~\rme^{-\tr a^2-S_{\mathrm{int}}(a)}~.
\label{Zis}
\end{equation}
Here we follow the so-called ``full Lie algebra'' approach \cite{Billo:2017glv,Fiol:2020bhf,Beccaria:2020hgy,Beccaria:2021hvt,Fiol:2021icm} and integrate over all elements
$a^b$ of the matrix with a measure given by
\begin{equation}
da=\prod_{b=1}^{N^2-1}\frac{da^b}{\sqrt{2\pi}}
\label{measure}
\end{equation}
in such a way that the Gaussian integration is normalized to 1. In (\ref{Zis}) $S_{\mathrm{int}}(a)$ represents an interaction term whose explicit form depends on the representation $\mathcal{R}$ in which the matter hypermultiplets transform.
As shown in
\cite{Billo:2019fbi,Beccaria:2020hgy}, for a generic $\mathcal{N}=2$ SU($N$) theory $S_{\mathrm{int}}(a)$ can be written as a linear combination of the following traces\,%
\footnote{We neglect the instanton contributions since we are ultimately interested in studying the 't Hooft large-$N$ limit where instantons are exponentially suppressed.}
\begin{equation}
\Tr_{\mathcal{R}}a^{2k}-\Tr_{\mathrm{adjoint}}a^{2k}~,
\label{traceR}
\end{equation}
which, when expressed in terms of the traces in the fundamental representation, become a superposition of double traces of the form $\big(\tr a^{\ell}\tr a^{2k-\ell}\big)$ and of single traces of the form $\tr a^{2k}$. From (\ref{traceR}), we see that 
if $\mathcal{R}$ is the adjoint representation, which is the case of the $\mathcal{N}=4$ SYM theory, then $S_{\mathrm{int}}(a)$ vanishes and the matrix model becomes free
with a purely Gaussian term. In a genuinely $\mathcal{N}=2$ theory the interacting part
$S_{\mathrm{int}}(a)$ is instead 
not zero and can be regarded as a deformation of the free Gaussian model. 

Given any function $f(a)$, its expectation value is defined as
\begin{equation}
\big\langle f(a)\big\rangle=\frac{1}{\mathcal{Z}}\,\int da~f(a)~\rme^{-\tr a^2-S_{\mathrm{int}}(a)} = \frac{\big\langle f(a)~
\rme^{-S_{\mathrm{int}}(a)} \big\rangle_0\phantom{\Big|}}{\big\langle\rme^{-S_{\mathrm{int}}(a)}\big\rangle_0\phantom{\Big|}}
\label{vev}
\end{equation}
where the notation $\langle~\rangle_0$ stands for the expectation value in the free
model\,%
\footnote{Here we have followed the conventions of \cite{Billo:2017glv,Beccaria:2020hgy,Beccaria:2021hvt} and instead of writing the Gaussian term as
$\rme^{-\frac{8\pi^2 N}{\lambda}\tr a^2}$, we have performed the rescaling $a\to \sqrt{\lambda/(8\pi^2N)}\,a$ to bring it to the canonical form. In this way all dependence on the coupling is inside the interaction action 
$S_{\mathrm{int}}(a)$ (see Eq.\,(\ref{Sint}) below).}. 
Through this formula, the calculation of any expectation value in the $\mathcal{N}=2$ matrix model is thus reduced to the calculation of expectation values 
in the free theory. Given the structure of $S_{\mathrm{int}}(a)$ that we have recalled above, which implies that $S_{\mathrm{int}}(-a)=S_{\mathrm{int}}(a)$, only functions that are even under the exchange $a\to-a$ may have a non-vanishing expectation value.

A natural set of operators to consider in the matrix model is that of the multi-traces
\begin{equation}
\Omega_{\mathbf{n}}=\tr a^{n_1}\,\tr a^{n_2}\ldots\tr a^{n_\ell}
\label{Omegan}
\end{equation}
which clearly obey the relation
\begin{equation}
\Omega_{\mathbf{n}}\,\Omega_{\mathbf{m}}=\Omega_{\mathbf{n},\mathbf{m}}~.
\end{equation}
We denote their expectation values as $T_{\mathbf{n}}$, namely
\begin{equation}
T_{\mathbf{n}}\,\equiv\,T_{n_1,\ldots,n_\ell}=\big\langle \tr a^{n_1}\,\tr a^{n_2}\ldots
\tr a^{n_\ell} \big\rangle~,
\label{Tn}
\end{equation}
that are non-zero only if $|\mathbf{n}|$ is even.

The operators $\Omega_{\mathbf{n}}$, however, are not the representatives 
in the matrix model
of the local chiral operators $\mathcal{O}_{\mathbf{n}}(x)$ of the gauge theory 
\cite{Gerchkovitz:2016gxx}. Indeed, the latter correspond to the normal-ordered version 
of $\Omega_{\mathbf{n}}$ which is defined by the Gram-Schmidt 
orthogonalization procedure:
\begin{equation}
\mathcal{O}_{\mathbf{n}}=\Omega_{\mathbf{n}}-
\sum_{|\mathbf{m}|<|\mathbf{n}|} \mathcal{C}_{\mathbf{n}}^{\,\mathbf{m}}\,\mathcal{O}_{\mathbf{m}}~.
\label{Onis}
\end{equation}
Here the mixing coefficients $\mathcal{C}_{\mathbf{n}}^{\,\mathbf{m}}$ are fixed by demanding that $\mathcal{O}_{\mathbf{n}}$ be orthogonal to all operators 
$\mathcal{O}_{\mathbf{m}}$ of lower dimensions, {\it{i.e.}}
\begin{equation}
\big\langle \mathcal{O}_{\mathbf{n}}\,\mathcal{O}_{\mathbf{m}} \big\rangle =
0\quad\mbox{for all}~|\mathbf{m}|<|\mathbf{n}|~.
\label{OnOm}
\end{equation}
Notice that 2-point functions $\big\langle
\cO_{\mathbf{n}} \,\cO_{\mathbf{m}}\big\rangle$ with 
$|\mathbf{n}|=|\mathbf{m}|$ are 
not required to be diagonal. Of course, it would be possible to redefine the operators 
and orthogonalize them as well. However, the same situation holds for the 
corresponding multi-trace operators in the gauge theory and there one does not usually redefine the operators by mixing different trace structures with the same dimension. Therefore here, as in most of the literature, we do not make this step and perform a Gram Schmidt procedure which is not complete.

Enforcing the condition (\ref{OnOm}) one can determine the mixing coefficients in terms of the expectation values (\ref{Tn}), and find that $\mathcal{C}_{\mathbf{n}}^{\,\mathbf{m}}$ is different from zero only when $|\mathbf{n}|$ and $|\mathbf{m}|$ are both even or both odd. For example, the double-trace operator $\mathcal{O}_{2,3}$ of dimension 5 can only mix with the single-trace operator $\mathcal{O}_3$ of dimension 3, and the mixing coefficient is
\begin{equation}
\mathcal{C}_{2,3}^{\phantom{2,3}3}=\frac{T_{2,3,3}}{T_{3,3}}~.
\end{equation}
Of course, when the dimension of the operators increases, the mixing coefficients
become more and more intricate. However, as shown in Appendix~\ref{app:mixing},
it is possible to write them in closed form as ratios of determinants of matrices constructed with the expectation values (\ref{Tn}).

The normal-ordered operators $\mathcal{O}_{\mathbf{n}}$ satisfy the relation
\begin{equation}
\mathcal{O}_{\mathbf{n}}\,\mathcal{O}_{\mathbf{m}}=
\mathcal{O}_{\mathbf{n},\mathbf{m}}+\ldots
\end{equation}
where the dots stand for terms of dimension smaller than $|\mathbf{n}|+|\mathbf{m}|$, which is the matrix-model counterpart of the chiral ring relation (\ref{chiralring}).
{From} the definition (\ref{Onis}) and the orthogonality condition (\ref{OnOm}), one
easily finds that
\begin{equation}
\big \langle \mathcal{O}_{\mathbf{n}} \,\mathcal{O}_{\mathbf{m}}\big\rangle
=\big \langle \Omega_{\mathbf{n}} \,\mathcal{O}_{\mathbf{m}}\big\rangle
=\big \langle \mathcal{O}_{\mathbf{n}} \,\Omega_{\mathbf{m}}\big\rangle
\quad\mbox{for}~|\mathbf{n}|=|\mathbf{m}|~.
\label{Onm1}
\end{equation}
Notice that the last term can be further manipulated and rewritten using
the expectation values (\ref{Tn}) as follows
\begin{equation}
\begin{aligned}
\big \langle \mathcal{O}_{\mathbf{n}}\,\Omega_{\mathbf{m}}\big\rangle
&=\big \langle \Omega_{\mathbf{n}} \,\Omega_{\mathbf{m}}\big\rangle-
\sum_{|\mathbf{p}|<|\mathbf{n}|}
\mathcal{C}_{\mathbf{n}}^{\,\mathbf{p}}\,\big \langle \Omega_{\mathbf{p}} \,
\Omega_{\mathbf{m}}\big\rangle
+\sum_{|\mathbf{q}|<|\mathbf{p}|<|\mathbf{n}|}
\mathcal{C}_{\mathbf{n}}^{\,\mathbf{p}}\,\mathcal{C}_{\mathbf{p}}^{\,\mathbf{q}}\,\big \langle 
\Omega_{\mathbf{q}}\,
\Omega_{\mathbf{m}}\big\rangle+\ldots\\[1mm]
&=T_{\mathbf{n},\mathbf{m}}-
\sum_{|\mathbf{p}|<|\mathbf{n}|}
\mathcal{C}_{\mathbf{n}}^{\,\mathbf{p}}\,T_{\mathbf{p},\mathbf{m}}
+\sum_{|\mathbf{q}|<|\mathbf{p}|<|\mathbf{n}|}
\mathcal{C}_{\mathbf{n}}^{\,\mathbf{p}}\,\mathcal{C}_{\mathbf{p}}^{\,\mathbf{q}}\,T_{\mathbf{q},\mathbf{m}}+\ldots~.
\end{aligned}
\label{Onm2}
\end{equation}
The various terms in the second line above can be summed and a closed-form expression for the correlator can be obtained as a ratio of determinants of matrices constructed
with the expectation values (\ref{Tn}) (see Appendix~\ref{app:mixing} for details).

As proposed in \cite{Baggio:2014sna,Baggio:2015vxa,Gerchkovitz:2016gxx,Baggio:2016skg}, the coefficient 
$\mathcal{G}_{\mathbf{n_1},\ldots,\,\mathbf{n_k};\mathbf{m}}$ appearing in the
extremal correlator (\ref{extremal}) of the gauge theory is entirely captured by the 2-point correlator between $\mathcal{O}_{\mathbf{n_1},\ldots,\,\mathbf{n_k}}$
and $\mathcal{O}_{\mathbf{m}}$ in the matrix model, namely
\begin{equation}
\mathcal{G}_{\mathbf{n_1},\ldots,\,\mathbf{n_k};\mathbf{m}}=
\big\langle \mathcal{O}_{\mathbf{n_1},\ldots,\,\mathbf{n_k}}
\,\mathcal{O}_{\mathbf{m}}\big\rangle~.
\end{equation}
{From} (\ref{OnOm}) and (\ref{Onm1}) it immediately follows that the right hand side  is non-vanishing only if 
$|\mathbf{n_1}|+\ldots+|\mathbf{n_k}|=|\mathbf{m}|$, in agreement with the selection rule
(\ref{selection}).
In case of single-trace operators these formulas simplify. In particular, for the 2-point correlators we have
$\mathcal{G}_{n;m}=G_n\,\delta_{n,m}$ with
\begin{equation}
G_n = \big\langle \mathcal{O}_n\,\mathcal{O}_n\big\rangle
=\big\langle \mathcal{O}_n \, \Omega_n \big\rangle~,
\label{Gn1}
\end{equation}
while for the 3-point correlators we have $\mathcal{G}_{n_1,n_2;m}=G_{n_1,n_2}\,\delta_{n_1+n_2,m}$ with
\begin{equation}
G_{n_1,n_2} = \big\langle \mathcal{O}_{n_1,n_2}\,\mathcal{O}_{n_1+n_2}\big\rangle =
\big\langle \Omega_{n_1,n_2}\,\mathcal{O}_{n_1+n_2}\big\rangle  =
\big\langle \mathcal{O}_{n_1,n_2}\,\Omega_{n_1+n_2}\big\rangle ~.
\label{Gn1n21}
\end{equation}
Thus, the calculation of the 2- and 3-point functions of the single-trace primary operators in the gauge theory is reduced to the calculation of the 2-point correlators in the matrix model and ultimately to the evaluation of the expectation values $T_{n_1,\ldots,n_k}$.

\subsection{Single-trace correlators at large \texorpdfstring{$N$}{}}
Several significant simplifications occur in the procedure outlined above, when one considers the 't Hooft large-$N$ limit in which $N\to\infty$ with
\begin{equation}
\lambda=Ng^2
\label{lambda}
\end{equation}
kept fixed. Focusing for simplicity on the single-trace observables, one can show that in the planar limit instead of the 
operators $\mathcal{O}_n$, which are orthogonal to {\emph{all}} operators of dimension smaller than $n$ with both single and multi-traces, it is enough to consider a set of simpler operators $O_n$, which are orthogonal to only the single-trace operators of dimension smaller than $n$. These operators are defined by
\begin{equation}
O_n=\Omega_n-\sum_{m<n}C_{n,m}\,O_m
\label{newOn}
\end{equation}
where the mixing coefficients $C_{n,m}$ are obtained by requiring that
\begin{equation}
\big\langle O_n\,O_m\big\rangle=0\quad\mbox{for}~m<n~.
\label{ortho1}
\end{equation}
This amounts to implement the Gram-Schmidt orthogonalization procedure only in the subspace of the single-trace operators. Thus, the operators $\mathcal{O}_n$ and $O_n$ differ from each other\,%
\footnote{Actually, $\mathcal{O}_n=O_n$ for $n=2,3,4,5$ since in these cases the mixing can occur only with single-trace operators. For $n>5$, instead, where also multi-trace operators appear, one has $\mathcal{O}_n\not=O_n$.}. However, one can show (see for example \cite{Baggio:2016skg,Beccaria:2020hgy}) that the difference is sub-leading in the large-$N$ expansion, {\it{i.e.}}
\begin{equation}
\mathcal{O}_n=O_n+\frac{1}{N}\big(\mbox{single and multi traces}\big)
\label{OOn}
\end{equation}
where the second term in the right hand side stands for single and multi-trace 
operators of dimension smaller than $n$ which, when inserted inside correlators,
yield contributions that are suppressed in the large-$N$ limit. It is worth pointing out that the
coefficients $\mathcal{C}_n^{\,m}$ and $C_{n,m}$, which account for the mixing of
the single-trace operator of dimension $n$ with the single-trace operator of dimension $m$ in the two schemes, are not the same but they agree at large $N$:
\begin{equation}
\mathcal{C}_n^{\,m} =C_{n,m}+O\big(1/N\big)~.
\label{CC}
\end{equation}

Using these properties, we can simplify the calculation of the 2- and 3-point correlators in the large-$N$ limit. Let us first consider the 2-point correlator (\ref{Gn1}), which upon using (\ref{newOn}) becomes
\begin{equation}
G_n=\big\langle \mathcal{O}_n \, \Omega_n \big\rangle=\big\langle \mathcal{O}_n \, O_n \big\rangle~.
\end{equation}
Then, exploiting (\ref{OOn}), we can replace $\mathcal{O}_n$ with $O_n$ and conclude that 
\begin{equation}
G_n=\big\langle O_n \, O_n \big\rangle+O\big(1/N\big)~.
\label{Gn2}
\end{equation}
This is precisely the form of the 2-point correlator that was used, for example, in the calculations reported in \cite{Beccaria:2020hgy,Beccaria:2021hvt}.

Also the 3-point correlators of single-trace operators in the large-$N$ limit can be written only 
in terms of the operators $O_n$. To show this, we first observe that in last term of (\ref{Gn1n21}) we
can substitute $\Omega_{n_1+n_2}$ with $O_{n_1+n_2}$
since the difference consists of operators of dimension smaller than $(n_1+n_2)$ which are orthogonal to $\mathcal{O}_{n_1,n_2}$. Thus we have
\begin{equation}
G_{n_1,n_2}=\big\langle \mathcal{O}_{n_1,n_2}\,\Omega_{n_1+n_2}\big\rangle=
\big\langle \mathcal{O}_{n_1,n_2}\,O_{n_1+n_2}\big\rangle~.
\label{Gn1n22}
\end{equation}
Then, we can use the following relation, proven in \cite{Baggio:2016skg}, 
\begin{equation}
\mathcal{O}_{n_1,n_2}=O_{n_1} O_{n_2}+\big(\mbox{single traces}\big)+\frac{1}{N}
\big(\mbox{multi traces}\big)~,
\label{relation}
\end{equation}
and upon substituting it in (\ref{Gn1n22}), we get
\begin{equation}
G_{n_1,n_2}=\big\langle O_{n_1}\,O_{n_2}\,O_{n_1+n_2}\big\rangle+O\big(1/N\big)~.
\label{Gn1n23}
\end{equation}
Indeed, the single traces in (\ref{relation}) are of dimension smaller than $(n_1+n_2)$ and, hence, are orthogonal to $O_{n_1+n_2}$, while the multi traces give rise to contributions which are suppressed when $N\to\infty$.

The 2- and 3-point correlators (\ref{Gn2}) and (\ref{Gn1n23}) can be explicitly evaluated with a moderate computational effort, and thus provide a very efficient way to obtain information on the 2- and 3-point correlation functions of the SYM theory in the planar limit.

\section{Two- and three-point functions in the \texorpdfstring{$\mathbf{E}$}{}
theory at large \texorpdfstring{$N$}{}: perturbative results}
\label{secn:Etheory}

In the following we will provide explicit examples of the functions $G_n$ and $G_{n_1,n_2}$ in the so-called $\mathbf{E}$ theory \cite{Billo:2019fbi,Beccaria:2020hgy}.
When the hypermultiplets are in the symmetric plus anti-symmetric representation of SU($N$), using (\ref{traceR}), one can show that the interacting part of the matrix model 
is \cite{Billo:2019fbi,Beccaria:2020hgy,Beccaria:2021hvt} 
\begin{equation}
S_{\mathrm{int}}(a)=4\sum_{\ell,m=1}^\infty
(-1)^{\ell+m}\Big(\frac{\lambda}{8\pi^2 N}\Big)^{\ell+m+1}\,
\frac{(2\ell+2m+1)!\,\zeta(2\ell+2m+1)}{(2\ell+1)!\,(2m+1)!}
\tr a^{2\ell+1}\,\tr a^{2m+1}
\label{Sint}
\end{equation}
where $\zeta$ are the Riemann zeta-values. The fact that, differently from what happens in other $\mathcal{N}=2$ superconformal theories, only products of two traces 
of odd powers of $a$ appear in (\ref{Sint}) is the main reason behind the possibility of
obtaining closed-form expressions and eventually extrapolate the perturbative results at strong coupling.

Using the interaction action (\ref{Sint}), it is rather straightforward to obtain the first perturbative contributions to the 2- and 3-point coefficients $G_n$ and $G_{n_1,n_2}$, even if the calculation becomes longer
and longer as the dimensions of the operators grow. This is clearly due to the fact 
that, even in the simplified set-up of the large-$N$ limit, an increasing number of mixing coefficients have to be determined in order to find the explicit expression of the
operators $O_n$ to be used in the correlators. Nevertheless, in the $\mathbf{E}$ theory 
the very first perturbative terms in $G_n$ and $G_{n_1,n_2}$ can be easily 
obtained in full generality at large $N$, as we are going to show in the following.

\subsection{Two-point functions}
The planar limit of the 2-point functions $G_n$ in the $\mathbf{E}$ theory has been extensively studied in \cite{Beccaria:2020hgy,Beccaria:2021hvt}, and here we simply recall the main results.
When $N\to \infty$, one finds\,%
\footnote{In the $\mathbf{E}$ theory, the first non-planar corrections are of order $1/N^2$, differently from other superconformal theories in which they are of order $1/N$.
In the matrix model this property is a consequence of the fact that the interaction action $S_{\mathrm{int}}(a)$ of the $\mathbf{E}$ theory contains only products of two odd traces, as one can see from (\ref{Sint}). On the contrary, the
interaction action of other superconformal theories contains also products of two even traces and/or terms with single traces. As  explicitly shown in \cite{Beccaria:2020hgy}, this different structure is responsible for the different behavior in the large-$N$ limit.}
\begin{equation}
G_n=G_n^{(0)}\Big[\big(1+\Delta_n\big)+O\big(1/N^2\big)\Big]~.
\label{Gnpert}
\end{equation}
Here $G_{n}^{(0)}$ denotes the 2-point function coefficient in the $\mathcal{N}=4$
SYM at large $N$, which in our normalization is given by
\begin{equation}
G_n^{(0)}=n\,\Big(\frac{N}{2}\Big)^n~,
\label{Gn0}
\end{equation} 
and $\Delta_n$ is a function of $\lambda$ representing the deviation from the $\mathcal{N}=4$ result. As shown in \cite{Beccaria:2020hgy,Beccaria:2021hvt},
when $n=2k$ 
\begin{equation}
\Delta_{2k}=0~,
\label{Deltaeven}
\end{equation}
and when $n=2k+1$ 
\begin{equation}
\Delta_{2k+1}=-\frac{\zeta(4k+1)}{2^{2k-1}}\,\binom{4k+2}{2k+1}\,\Big(\frac{\lambda}{8\pi^2}\Big)^{2k+1}+O\big(\lambda^{2k+2}\big)~.
\label{Deltaoddpert}
\end{equation}
Thus, in the large-$N$ limit the 2-point functions of operators of even dimensions 
of the $\mathbf{E}$ theory coincide with those of the $\mathcal{N}=4$ SYM, while
those of operators of odd dimension are different. 

\subsection{Three-point functions}
So far, the 3-point functions $G_{n_1,n_2}$ in the $\mathbf{E}$ theory have not been studied. Here we report the results of the calculations in the large-$N$ limit that we have performed following the procedure described above (for details we refer again to \cite{Beccaria:2020hgy,Beccaria:2021hvt}). Analyzing in detail numerous examples, we find that the general structure of the 3-point functions is
\begin{equation}
G_{n_1,n_2}=G_{n_1,n_2}^{(0)}\Big[\big(1+\Delta_{n_1,n_2}\big)+O\big(1/N^2\big)\Big]
\label{Gn1n2pert}
\end{equation}
where
\begin{equation}
G_{n_1,n_2}^{(0)}=\frac{n_1n_2(n_1+n_2)}{2}\,\Big(\frac{N}{2}\Big)^{n_1+n_2-1}
\label{Gn1n20}
\end{equation}
is the 3-point coefficient in the $\mathcal{N}=4$
SYM at large $N$, and $\Delta_{n_1,n_2}$ is a function of $\lambda$. 
When $n_1=2k$ and $n_2=2\ell$ we get
\begin{equation}
\Delta_{2k,2\ell}=0
\label{Deltaeveneven}
\end{equation}
in analogy with (\ref{Deltaeven}). Thus, like the 2-point functions, also the 3-point functions of operators of even dimension do not deviate from those of the $\mathcal{N}=4$ theory at large $N$. When $n_1=2k$ and $n_2=2\ell+1$ we find
\begin{equation}
\Delta_{2k,2\ell+1}=-\frac{\zeta(4\ell+1)}{2^{2\ell-1}}\,\binom{4\ell+2}{2\ell+1}\,\Big(\frac{\lambda}{8\pi^2}\Big)^{2\ell+1}+O\big(\lambda^{2\ell+2}\big)~.
\label{Deltaevenodd}
\end{equation}
This is exactly the first perturbative contribution of order $\lambda^{2\ell+1}$
to $\Delta_{2\ell+1}$ as we see from (\ref{Deltaoddpert}). Thus, we can write
\begin{equation}
\Delta_{2k,2\ell+1}=\Delta_{2\ell+1}\big|_{\lambda^{2\ell+1}}+O\big(\lambda^{2\ell+2}\big)~.
\label{Deltaevenodd1}
\end{equation}
Finally, when $n_1=2k+1$ and $n_2=2\ell+1$, with $k\leq \ell$, we have
\begin{equation}
\Delta_{2k+1,2\ell+1}=\big(1+\delta_{k,\ell}\big)
\Delta_{2k+1}\big|_{\lambda^{2k+1}}+O\big(\lambda^{2k+2}\big)~.
\label{Deltaoddodd1}
\end{equation}
These results show that, like the 2-point functions, also the 3-point functions involving operators 
of odd dimensions are different from those of the $\mathcal{N}=4$ theory even 
in the large-$N$ limit\,%
\footnote{One can easily check that the tree-level 2- and 3-point functions given in (\ref{Gn0}) and (\ref{Gn1n20}) are in agreement with the chiral ring relation (\ref{chiralring}). Furthermore, they are related as follows: $G_{n_1,n_2}^{(0)}=\frac{n_1+n_2}{N}\,
G_{n_1}^{(0)}\, G_{n_2}^{(0)}$. Using (\ref{Deltaeveneven}), (\ref{Deltaevenodd1}) and (\ref{Deltaoddodd1}), it is easy to realize that the same relation holds also in the interacting theory at the first perturbative order.}.

\subsection{Diagrammatic interpretation}
\begin{figure}[H]
\center{\includegraphics[scale=0.25]{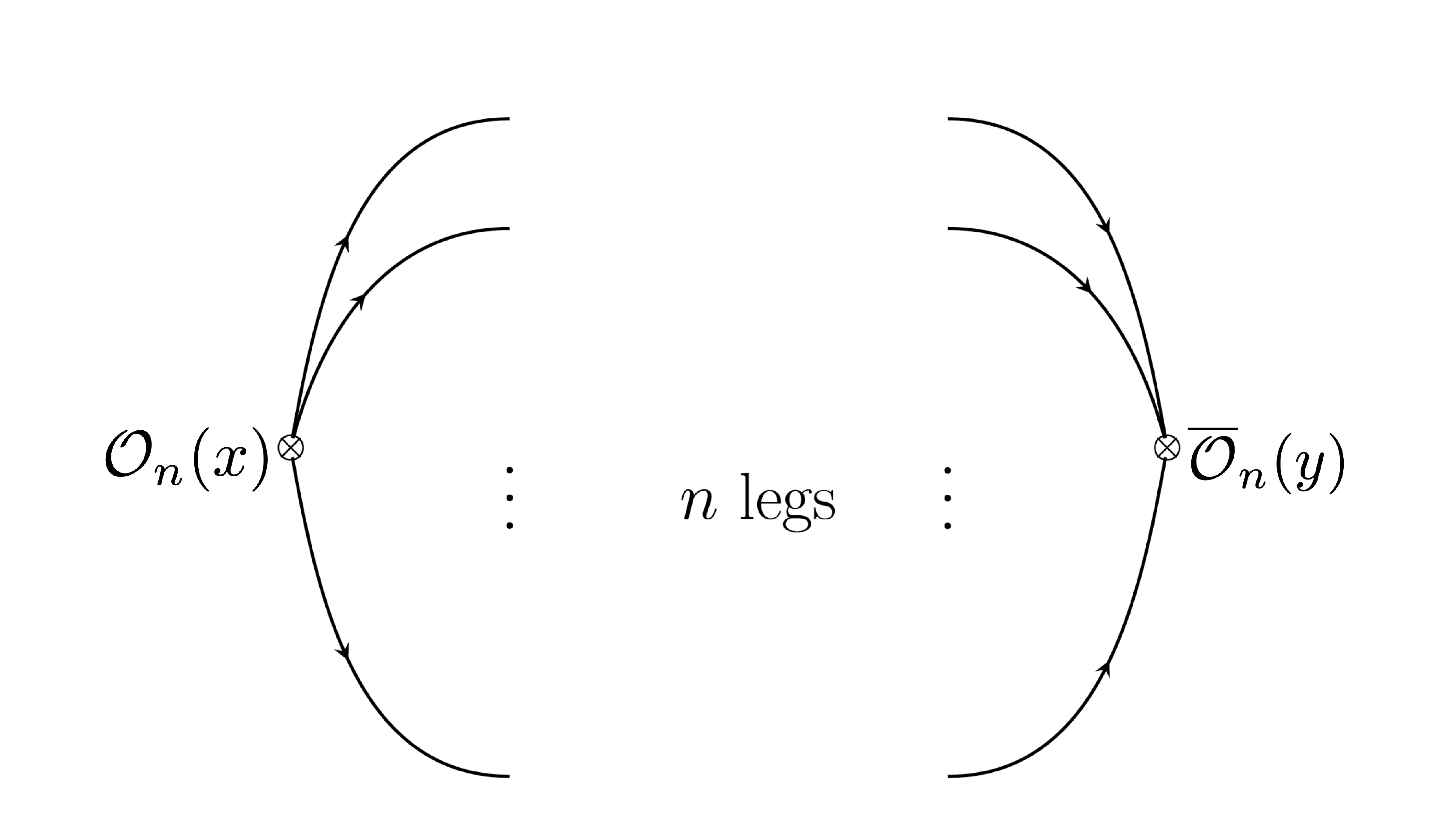}
\caption{Graphical representation of the chiral and anti-chiral operators $\mathcal{O}_{n}(x)$ and $\overbar{\mathcal{O}}_n(y)$. The outgoing lines represent the chiral field $\varphi$ while the incoming lines represent the anti-chiral field $\overbar{\varphi}$. \label{fig:1}} 
    }  
\end{figure}

The results reported above have a nice diagrammatic interpretation. 
Indeed, the 2-point function $G_n$ corresponds to the diagram in Fig.~\ref{fig:1},
where the $n$ legs of $\mathcal{O}_{n}(x)$ have 
to be contracted with the $n$ legs of $\overbar{\mathcal{O}}_{n}(y)$ using the rules and methods explained in \cite{Billo:2019fbi,Beccaria:2020hgy}.

\begin{figure}[ht]
\center{\includegraphics[scale=0.24]{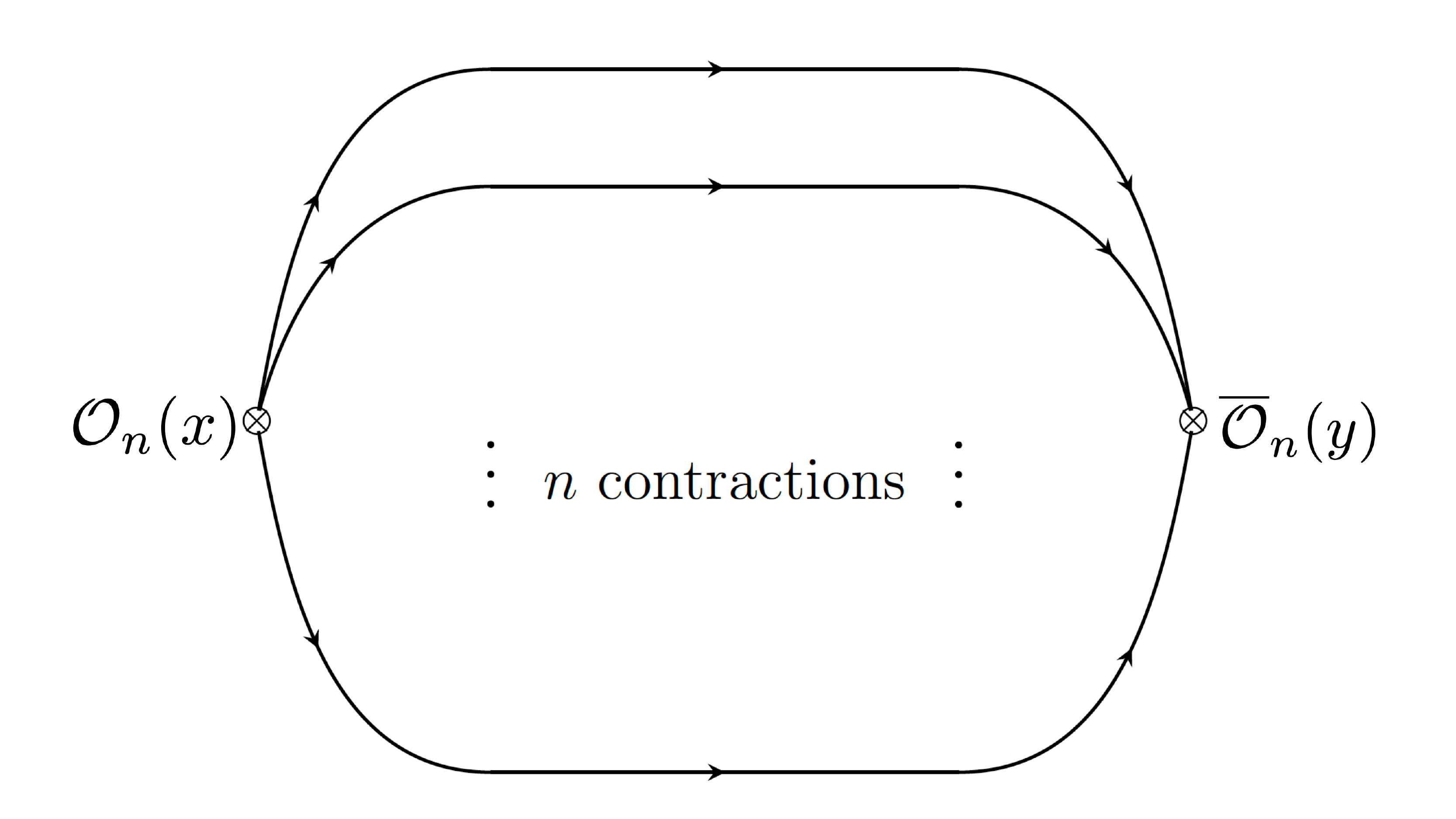}
\caption{When the contractions are made with free propagators one obtaines the tree-level result $G_n^{(0)}$ at the leading order for $N\to\infty$ is given by (\ref{Gn0}). \label{fig:2}} 
    }  
\end{figure}

\begin{figure}[ht]
\center{\includegraphics[scale=0.33]{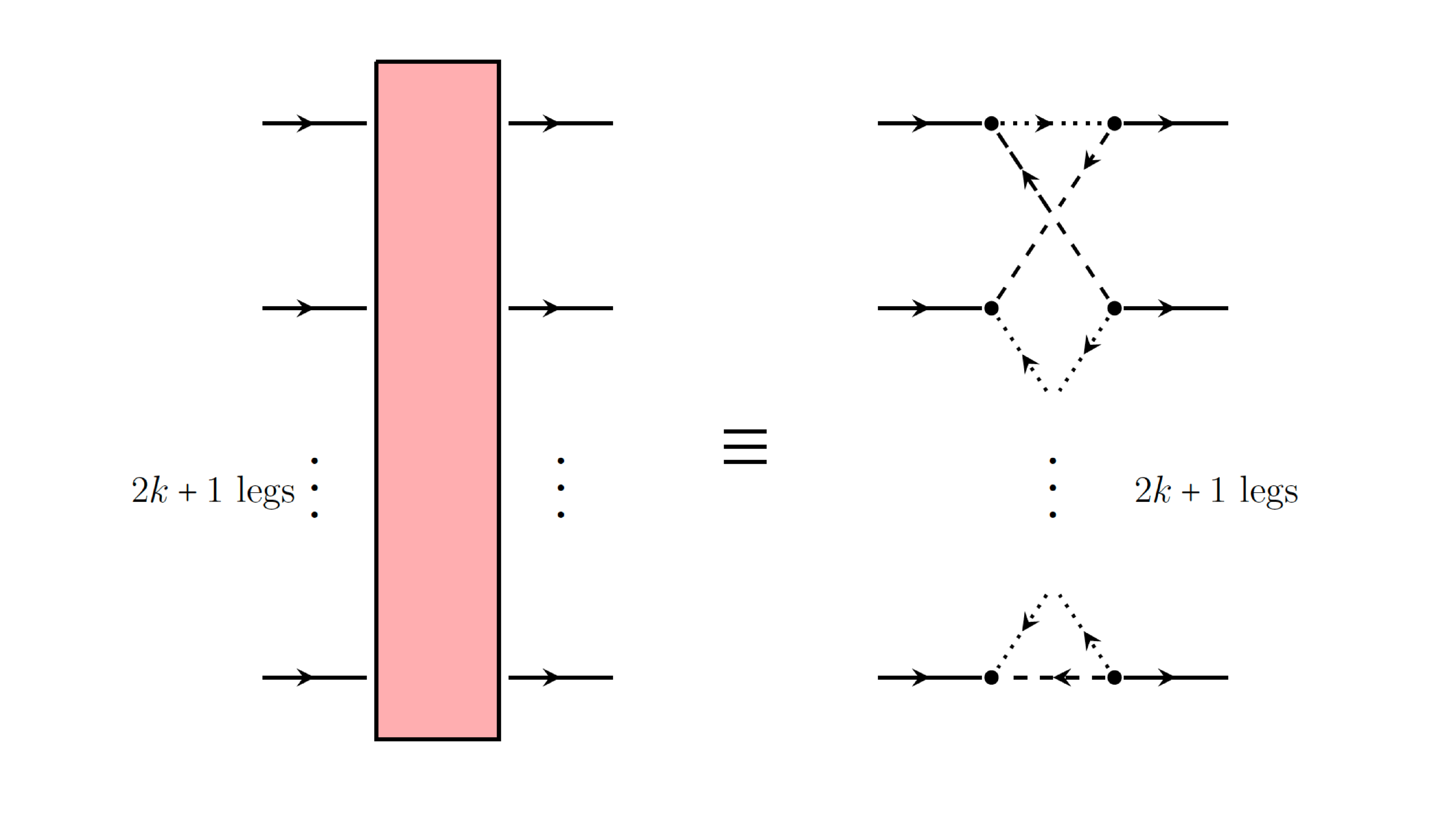}
\caption{The diagram which produces a planar contribution when inserted in the contraction of $(2k+1)$ legs as an effective vertex. In the right hand side we use the same notation of \cite{Billo:2017glv,Billo:2019fbi} for the matter hypermultiplets represented by the dashed and dotted lines. 
This contribution turns out to be proportional to $\zeta(4k+1)\,\lambda^{2k+1}$, as a result of the integration over the loop momenta according to \cite{Usyukina:1993ch}. \label{fig:3}} 
    }  
\end{figure}

\begin{figure}[H]
\center{\includegraphics[scale=0.28]{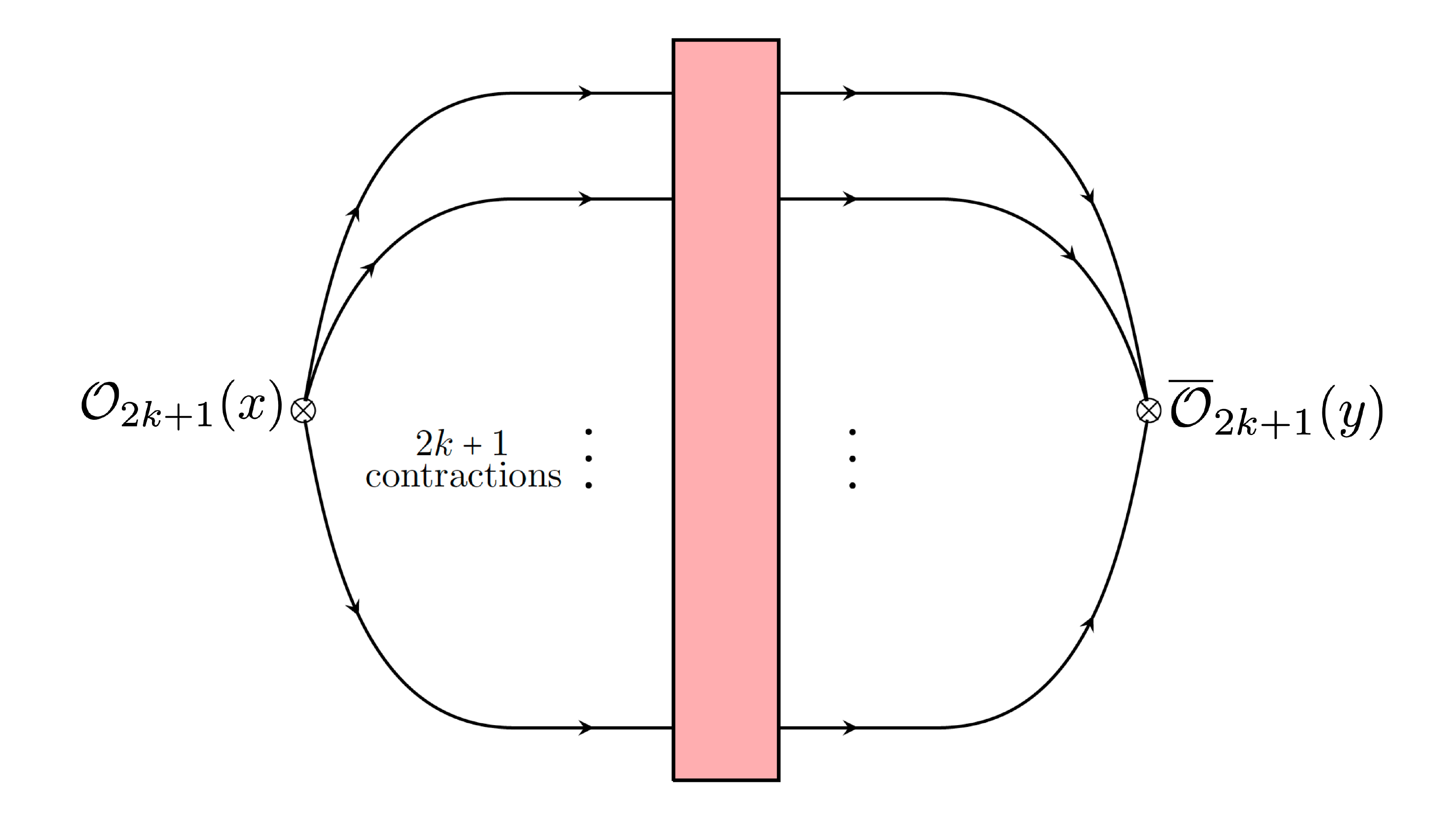}
\caption{When the effective vertex of Fig.~\ref{fig:3} is used in the contraction of the $(2k+1)$ legs, one gets the first perturbative contribution to the 2-point function $G_{2k+1}$ proportional to $\zeta(4k+1)\,\lambda^{2k+1}$, as given in (\ref{Deltaoddpert}). \label{fig:4}} 
    }  
\end{figure}

If these contractions are made with free propagators like in Fig.~\ref{fig:2}, one obtains the tree-level contribution $G_{n}^{(0)}$ whose leading term at large $N$ is given by (\ref{Gn0}).
If instead the contractions are made with the insertion of interaction vertices, one gets the loop corrections. In \cite{Beccaria:2020hgy} it was proved that when $n=2k$ there is no planar diagram that contributes to the 2-point function so that $\Delta_{2k}=0$, in agreement with the matrix model result\,%
\footnote{This argument is based on the fact that in the 2-point functions of even operators the first perturbative contribution is produced by a structure similar to that represented in
Fig. \ref{fig:3}, but with an even number of incoming and outgoing legs. A careful analysis of the color factor associated to this structure shows that this contribution either vanishes or is sub-leading in $N$ with respect to the tree-level term. For details we refer to Section\,7.1 of \cite{Beccaria:2020hgy}.
\label{footnote9}}. On the contrary, when $n=2k+1$ one can show that the first perturbative contribution at large $N$ arises from the diagram represented in Fig.~\ref{fig:3},
which is proportional to $\zeta(4k+1)\,\lambda^{2k+1}$. 
The insertion of this structure in the 
contraction between $\mathcal{O}_{2k+1}(x)$ and $\overbar{\mathcal{O}}_{2k+1}(y)$, as 
shown in Fig.~\ref{fig:4}, yields precisely the term proportional to $\lambda^{2k+1}$ in $G_{2k+1}$,
namely $G_{2k+1}^{(0)}\,\Delta_{2k+1}\big|_{\lambda^{2k+1}}$, in agreement with the matrix model calculation.

\begin{figure}[ht]
\center{\includegraphics[scale=0.30]{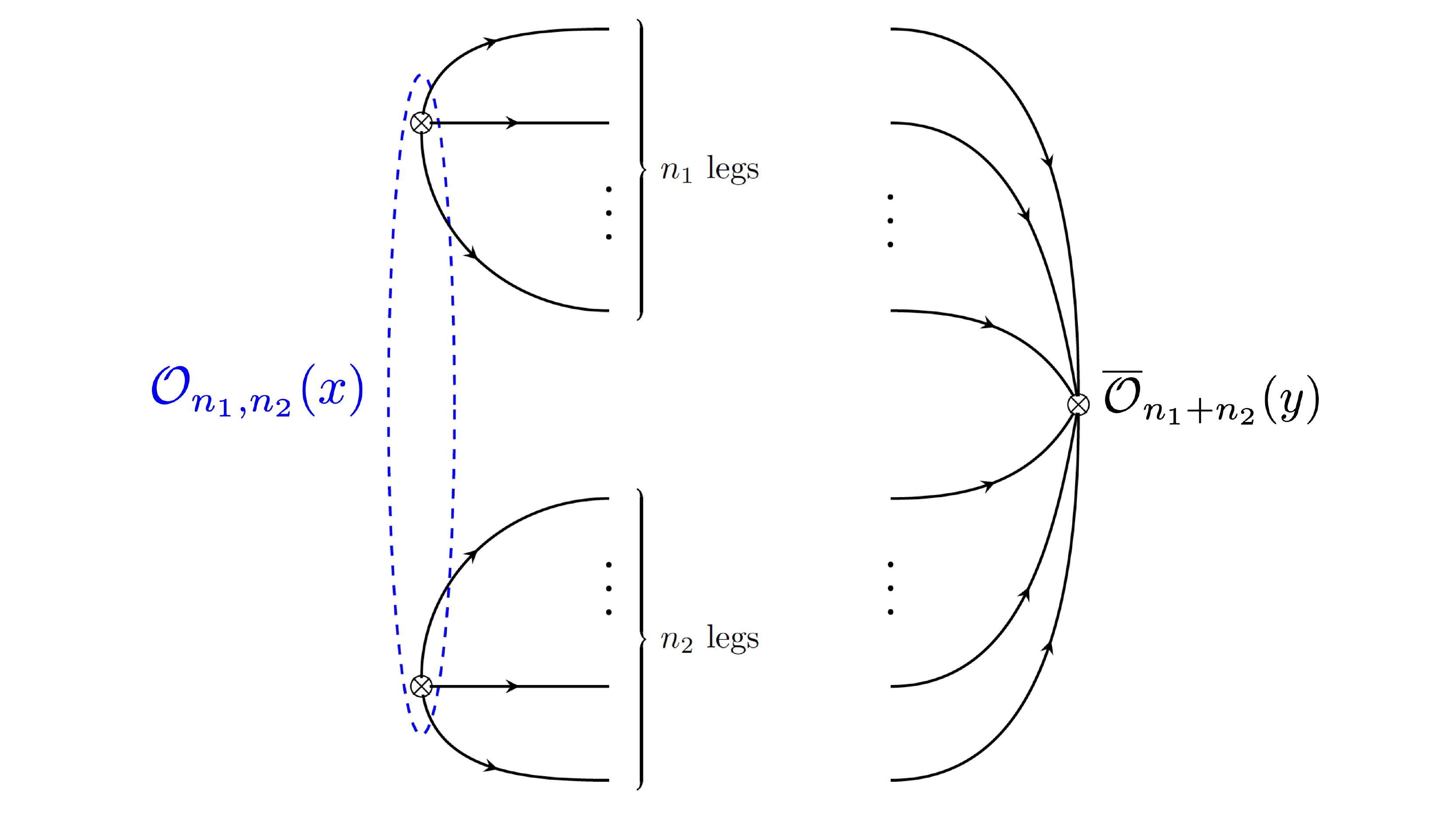}
\caption{Graphical representation of the chiral operator $\mathcal{O}_{n_1,n_2}(x)$ on the left, with two groups of legs corresponding to its two traces, and of the anti-chiral operator $\overbar{\mathcal{O}}_{n_1+n_2}(y)$ on the right. \label{fig:5}} 
    }  
\end{figure}

This analysis can be easily extended to the 3-point function $G_{n_1,n_2}$ which is obtained from the diagram in Fig.~\ref{fig:5}. 

\begin{figure}[ht]
\center{\includegraphics[scale=0.34]{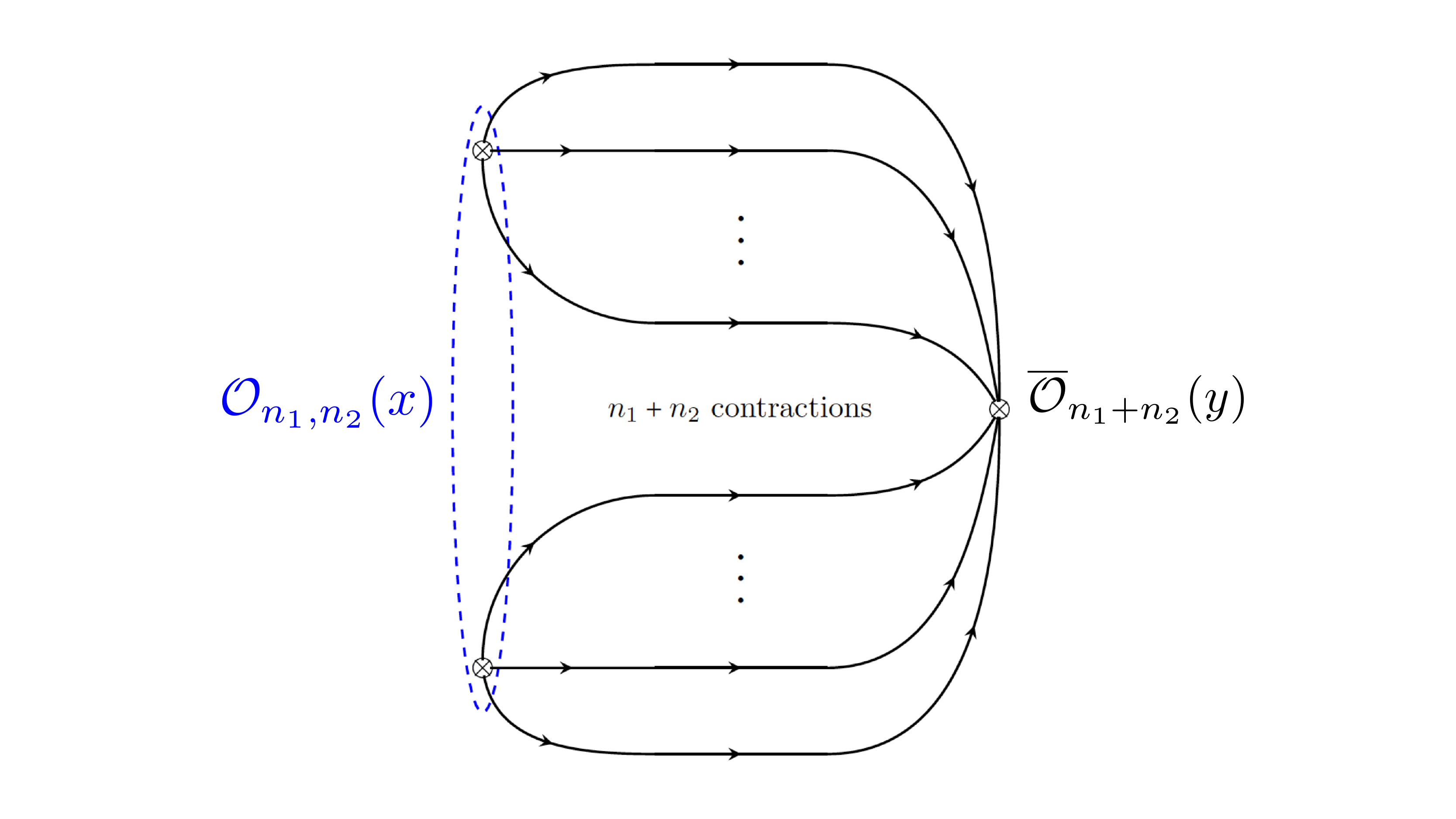}
\caption{By contracting all legs with the tree-level propagators one obtains the coefficient $G_{n_1,n_2}^{(0)}$ whose leading term at large $N$ is given in (\ref{Gn1n20}). \label{fig:6}} 
    }  
\end{figure}

\begin{figure}[ht]
\center{\includegraphics[scale=0.34]{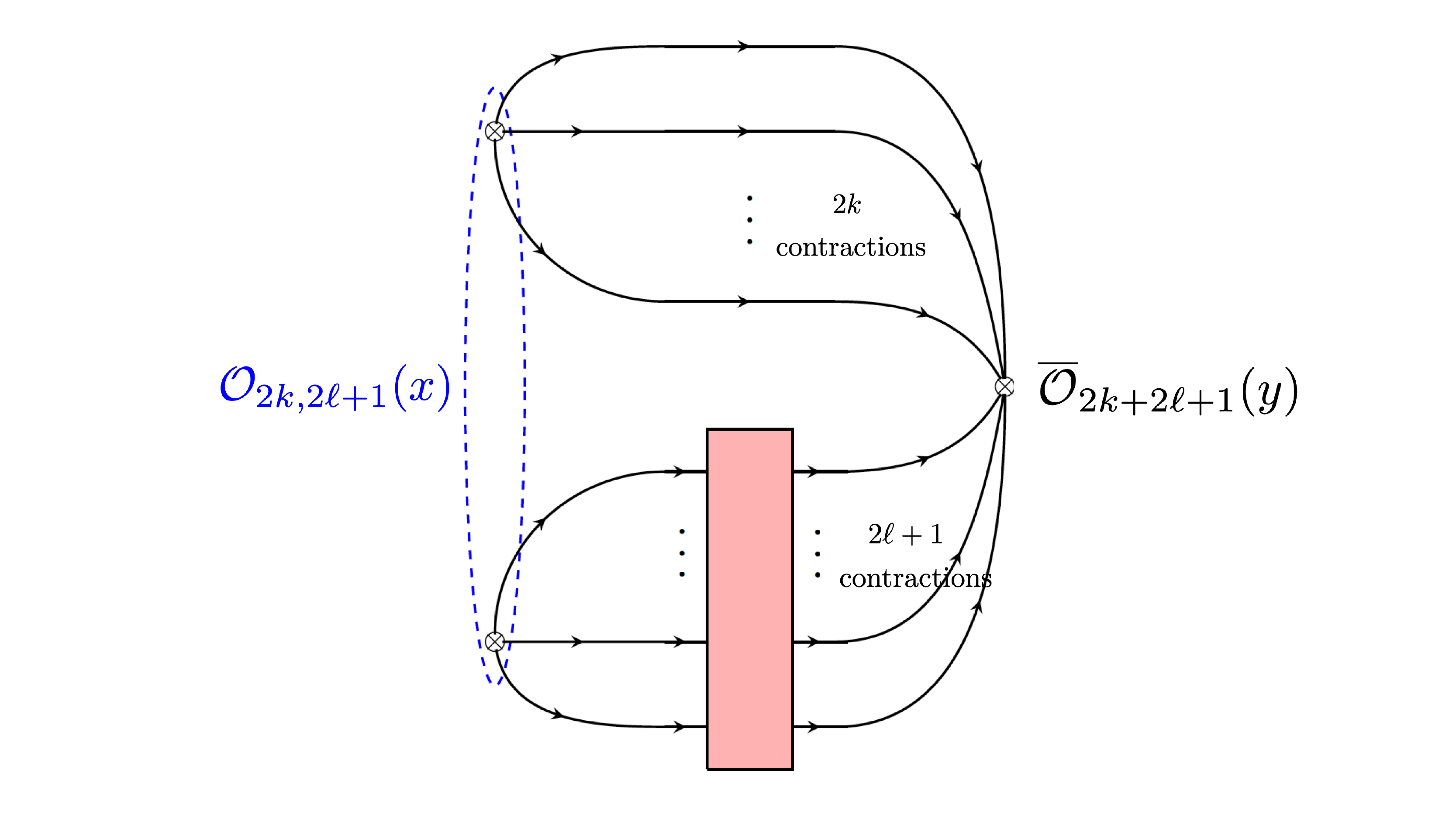}
\caption{The effective vertex of Fig.~\ref{fig:3} is used in the contraction of the $(2\ell+1)$ legs
and yields a planar contribution proportional to $\zeta(4\ell+1)\,\lambda^{2\ell+1}$ in the 3-point function $G_{2k,2\ell+1}$. \label{fig:7}} 
    }  
\end{figure}

By contracting the $(n_1+n_2)$ legs of $\mathcal{O}_{n_1,n_2}(x)$ with
those of $\overbar{\mathcal{O}}_{n_1+n_2}(y)$ with free propagators as shown in
Fig.~\ref{fig:6}, one obtains the tree-level result $G_{n_1,n_2}^{(0)}$ whose leading term at large $N$ is given by (\ref{Gn1n20}). 
Using the same argument of the 2-point functions of even operators 
(see footnote\,\ref{footnote9}), one can show that 
no loop diagram can contribute to the planar limit of $G_{2k,2\ell}$, so that 
$\Delta_{2k,2\ell}=0$ in agreement with the matrix model result. If instead either $n_1$ 
or $n_2$ or both are odd, there is a non-trivial planar contribution because we 
can insert the
structure of Fig.~\ref{fig:3} in the contraction. For example, if $n_1=2k$ and 
$n_2=2\ell+1$, we
can draw the diagram shown in Fig.~\ref{fig:7} which produces the term
$G_{2k,2\ell+1}^{(0)}\,\Delta_{2\ell+1}\big|_{\lambda^{2\ell+1}}$ in the planar limit, 
in full agreement with the matrix model calculation.
When $n_1=2k+1$ and $n_2=2\ell+1$, we can use the effective vertex of Fig.~\ref{fig:3} to contract either the $n_1$ legs or the $n_2$ legs of the two traces inside $\mathcal{O}_{n_1,n_2}(x)$
with those of $\overbar{\mathcal{O}}_{n_1+n_2}(y)$. Suppose that $k<\ell$. In this case the first perturbative correction arises
from the insertion of the effective vertex in the contraction of the $2k+1$ legs emanating from the first trace of $\mathcal{O}_{2k+1,2\ell+1}(x)$. This insertion produces a planar term proportional to $\zeta(4k+1)\,\lambda^{2k+1}$ in agreement with the matrix model result reported in (\ref{Deltaoddodd1}). 
The insertion of the effective vertex in the contraction of the $2\ell+1$ legs of the second trace of
$\mathcal{O}_{2k+1,2\ell+1}(x)$ also produces a planar term, but this is of higher order because
$\ell>k$. 
Finally, if $k=\ell$ the effective vertex can obviously be used in the contraction of the legs of
both traces, and this fact accounts for the factor of 2 which 
appears in (\ref{Deltaoddodd1}) when $k=\ell$.

\section{A simple three-point function at large \texorpdfstring{$N$}{}}
\label{secn:example}
Before addressing the strong-coupling behavior of the 3-point functions $G_{n_1,n_2}$ in full generality, we discuss in detail a simple example, namely $G_{2,3}$. 
In the matrix model this 3-point function is given by\,%
\footnote{Here and in the following, the symbol $\simeq$ means that only the leading term in the large-$N$ expansion is written.}
\begin{equation}
G_{2,3}\simeq\big\langle O_2\,O_3\,O_5\big\rangle
\label{G23}
\end{equation}
where, according to the definitions presented in Section~\ref{secn:gauge}, the operators are
\begin{equation}
O_2=\Omega_2-T_2~,~~O_3=\Omega_3~,~~O_5=\Omega_5-\frac{T_{3,5}}{T_{3,3}}\,\Omega_3~.
\label{O2O3O5}
\end{equation}
Using these expressions, we easily obtain
\begin{equation}
\big\langle O_2\,O_3\,O_5\big\rangle=
\big\langle \Omega_2\,O_3\,O_5\big\rangle=T_{2,3,5}-\frac{T_{3,5}}{T_{3,3}}\,T_{2,3,3}~.
\label{O2O3O5bis}
\end{equation}
The expectation values $T_{2,n_1,n_2}$ satisfy the relation (proven in Appendix~\ref{app:recursion}, see in particular (\ref{T2nisfin}))
\begin{equation}
T_{2,n_1,n_2}=\frac{1}{2}\big(N^2-1+n_1+n_2-2\lambda\,\partial_\lambda\mathcal{F}\big)\,T_{n_1,n_2}
+\lambda\,\partial_\lambda T_{n_1,n_2}
\label{recursionT2}
\end{equation}
where $\mathcal{F}=-\log\mathcal{Z}$ is the free energy. Exploiting this relation, after straightforward manipulations we can rewrite (\ref{O2O3O5bis}) as follows
\begin{equation}
\big\langle O_2\,O_3\,O_5\big\rangle=T_{3,5}+\lambda\,\partial_\lambda\Big(
\frac{T_{3,5}}{T_{3,3}}\Big)\,T_{3,3}~.
\label{O2O3O5ter}
\end{equation}
The crucial observation is that it is possible to write an exact formula for the expectation values $T_{n_1,n_2}$ in the $\mathbf{E}$ theory that is valid for all values of the 't Hooft coupling at large $N$. This formula makes use of the infinite matrix $\Xx$, firstly introduced in \cite{Beccaria:2020hgy,Beccaria:2021hvt}, which is related
to the partition function $\mathcal{Z}$ of the matrix model in the following way
\begin{equation}
\mathcal{Z}={\det}^{-\frac{1}{2}}\big(\mathbb{1}-\Xx\big)~,
\label{ZX}
\end{equation}
and whose elements are
\begin{equation}
\mathsf{X}_{k,\ell}=
 -8 (-1)^{k+\ell} \sqrt{(2k+1)(2\ell+1)} \int_0^\infty \!\frac{dt}{t}\, 
		\frac{\rme^t}{(\rme^t-1)^2}\,
		J_{2k+1}\Big(\frac{t\sqrt{\lambda}}{2\pi}\Big)\, 
		J_{2\ell+1}\Big(\frac{t\sqrt{\lambda}}{2\pi}\Big)
\label{Xkl}
\end{equation}
where $k,\ell\geq 1$ and $J$ are the Bessel functions of the first kind.
More precisely, in \cite{Beccaria:2021hvt} it was proved that at large $N$ the
even expectation values $T_{2k,2\ell}$ do not depend on $\lambda$ and 
thus coincide with those of the $\mathcal{N}=4$ SYM, namely
\begin{equation}
T_{2k,2\ell}\simeq\frac{N^{k+\ell+2}}{2^{k+\ell}}\,\frac{(2k)!\,(2\ell)!}{k!\,(k+1)!\,\ell!\,(\ell+1)!}~,
\label{Tkleven}
\end{equation}
while the odd expectation values $T_{2k+1,2\ell+1}$ depend in a non-trivial way on $\lambda$ and at large $N$ are given by
\begin{equation}
T_{2k+1,2\ell+1}\simeq\Big(\frac{N}{2}\Big)^{k+\ell+1}\sum_{i=0}^{k-1}\sum_{j=0}^{\ell-1}
c_{k,i}\,c_{\ell,j}\,\Big(\frac{1}{\mathbb{1}-\Xx}\Big)_{k-i,\ell-j}
\label{Tkl}
\end{equation}
where
\begin{equation}
c_{k,i}=\binom{2k+1}{i}\,\sqrt{2k-2i+1}~.
\label{cki}
\end{equation}
In particular, the correlators $T_{3,3}$ and $T_{3,5}$ appearing 
in (\ref{O2O3O5ter}) are
\begin{equation}
\begin{aligned}
T_{3,3}&\simeq\frac{3N^3}{8}\,\Big(\frac{1}{\mathbb{1}-\Xx}\Big)_{1,1}~,\\[2mm]
T_{3,5}&\simeq\frac{15N^4}{16}\,\bigg[\Big(\frac{1}{\mathbb{1}-\Xx}\Big)_{1,1}+
\frac{1}{\sqrt{15}}\,\Big(\frac{1}{\mathbb{1}-\Xx}\Big)_{1,2}\bigg]~.
\end{aligned}
\label{T35T33}
\end{equation}
Expanding the matrix $\Xx$ for small $\lambda$ by exploiting the well-known expansion of the Bessel functions, and using the result in the above expressions, it is possible to generate very long series in a quite efficient way. For example, the first few terms in the perturbative expansion of the correlator $\big\langle O_2\,O_3\,O_5\big\rangle$ obtained with this method are:
\begin{align}
\big\langle O_2\,O_3\,O_5\big\rangle&\simeq\frac{15N^4}{16}\bigg[
1-10\,\zeta(5)\Big(\frac{\lambda}{8\pi^2}\Big)^3+\frac{245\,\zeta(7)}{2}\Big(\frac{\lambda}{8\pi^2}\Big)^4-\frac{2331\,\zeta(9)}{2}\Big(\frac{\lambda}{8\pi^2}\Big)^5\notag\\
&\quad+4\big(2541\,\zeta(11)+25\,\zeta(5)^2\big)\Big(\frac{\lambda}{8\pi^2}\Big)^6
-\frac{13}{4}\big(26169\,\zeta(13)+700\,\zeta(5)\,\zeta(7)\big)\Big(\frac{\lambda}{8\pi^2}\Big)^7\notag\\
&\quad+\frac{105}{16}\big(105963\,\zeta(15)+3072\,\zeta(5)\,\zeta(9)+1988\,\zeta(7)^2\big)\Big(\frac{\lambda}{8\pi^2}\Big)^8+\ldots\bigg]~.
\end{align}
Actually we have generated all terms of $\big\langle O_2\,O_3\,O_5\big\rangle$
up to order $\lambda^{139}$ without any difficulty. These long expansions are very useful for the numerical analysis, as we will see in the next subsection.

More importantly, using the asymptotic expansion for large $\lambda$ of the Bessel functions appearing in (\ref{Xkl}) and then performing a Mellin transform, 
it is possible to obtain the strong-coupling behavior of the matrix 
$\Xx$ and show \cite{Beccaria:2021hvt} that when $\lambda\to\infty$ it becomes three-diagonal with elements given by
\begin{equation}
\Xx_{k,\ell}= (-1)^{k+\ell+1}\sqrt{\frac{2\ell+1}{2k+1}}\Big(\frac{\delta_{k-1,\ell}}{k(2k-1)}+\frac{\delta_{k,\ell}}{k(k+1)}+\frac{\delta_{k+1,\ell}}{(k+1)(2k+3)}\Big)\,\frac{\lambda}{8\pi^2}+O(\lambda^0)~.
\label{Xstrong}
\end{equation}
From this expansion, following the procedure explained in Appendix~A 
of \cite{Beccaria:2021hvt}, one can show that
\begin{equation}
\Big(\frac{1}{\mathbb{1}-\Xx}\Big)_{k,\ell}=\frac{4\pi^2}{\lambda}\,\sqrt{(2k+1)(2\ell+1)}\times \begin{cases}
~k(k+1)&~~\mbox{if}~k\leq \ell~,\\[1mm]
~\ell(\ell+1)&~~\mbox{if}~k\geq \ell~,
\end{cases}
\end{equation}
up to terms of order $1/\lambda^2$.
Using this result in (\ref{Tkl}) and performing the sums over $i$ and $j$, 
one finally gets 
\begin{equation}
T_{2k+1,2\ell+1}\simeq \,\frac{4\pi^2}{\lambda}\,\Big(\frac{N}{2}\Big)^{k+\ell+1}
\frac{(2k+1)!}{k!\,(k-1)!}\,\frac{(2\ell+1)!}{\ell!\,(\ell-1)!}\,\frac{1}{k+\ell}
+O\big(1/\lambda^2\big)
\,\equiv\,T_{2k+1,2\ell+1}^{(\infty)}~.
\label{Tklstrong}
\end{equation}

We have now all ingredients to write the correlator $\big\langle O_2\,O_3\,O_5\big\rangle$ at strong coupling. In fact, from (\ref{Tklstrong}) we see that for $\lambda\to\infty$
\begin{equation}
T_{3,3}\simeq \,T_{3,3}^{(\infty)}=\frac{9N^3\pi^2}{\lambda}
+O\big(1/\lambda^2\big)
\quad\mbox{and}\quad
T_{3,5}\simeq \,T_{3,5}^{(\infty)}=\frac{30N^4\pi^2}{\lambda} +
O\big(1/\lambda^2\big)~.
\label{T35T33strong}
\end{equation}
This implies that at leading order in the large-$\lambda$ expansion, the ratio
$T_{3,5}/T_{3,3}$ is constant so that the $\lambda$-derivative 
term in (\ref{O2O3O5ter}) can be discarded, leaving us with
\begin{equation}
\big\langle O_2\,O_3\,O_5\big\rangle\simeq
\frac{30N^4\pi^2}{\lambda}+O\big(1/\lambda^2\big)~.
\label{O2O3O5quater}
\end{equation}
Writing the result in the form (\ref{Gn1n2pert}), namely
\begin{equation}
G_{2,3}=\frac{15N^4}{16}\Big[\big(1+\Delta_{2,3}\big)+O\big(1/N^2\big)\Big]~,
\label{G23strong}
\end{equation}
we see from (\ref{O2O3O5quater}) that the strong-coupling expansion of the deviation 
$\Delta_{2,3}$ is given by
\begin{equation}
\Delta_{2,3}=-1+\frac{32\pi^2}{\lambda} +O\big(1/\lambda^2\big)~.
\label{Delta23strong}
\end{equation}

\subsection{Numerical checks}
\begin{figure}[t]
\center{\includegraphics[scale=0.32]{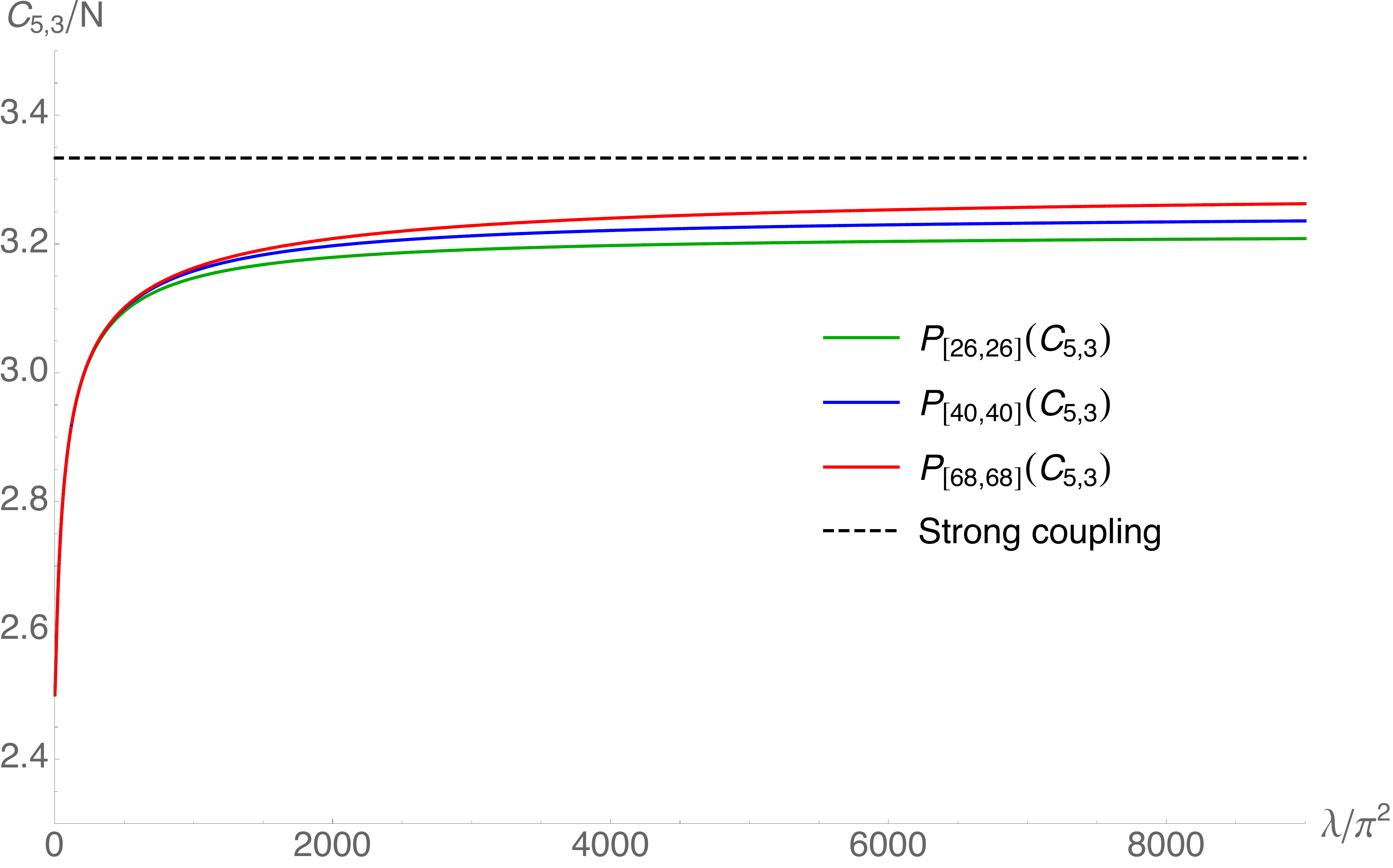}
\caption{Comparison between the Pad\'e curves $P_{[M/M]}(C_{5,3})$ for $M=26$ (green curve), $M=40$ (blue curve), $M=68$ (red curve) and the large-$\lambda$ theoretical prediction (\ref{C53strong}) (black dashed curve) for the mixing coefficient $C_{5,3}$ divided by $N$.\label{fig:8}} 
    }  
\end{figure}

Here provide a few numerical checks that we have performed in order to test the above results. First of all, using the perturbative expansion of the matrix $\Xx$ inherited from that of the Bessel functions appearing in (\ref{Xkl}), 
we have generated very long series for the expectation values
$T_{3,3}$ and $T_{3,5}$ and numerically evaluated the coefficients up to order $\lambda^{139}$. 
Using these expansions, we have then obtained the mixing coefficient of $O_5$ in the form
\begin{equation}
C_{5,3}=\frac{T_{3,5}}{T_{3,3}}=
\sum_{k=0}^{139} C_{5,3}^{(k)}\,\Big(\frac{\lambda}{\pi^2}\Big)^k + O\big(\lambda^{140}\big)
\label{C53pert}
\end{equation}
where
\begin{equation}
C_{5,3}^{(0)}\,\simeq\,\frac{5N}{2}
\label{C530}
\end{equation}
is the mixing coefficient in the free theory, {\it{i.e.}} in the $\mathcal{N}=4$ SYM.
The series (\ref{C53pert}) has a radius of convergence at $\lambda=\pi^2$ (see for example
the discussion in \cite{Beccaria:2020hgy,Beccaria:2021hvt}) but it can be extended beyond this bound with a Pad\'e resummation. Therefore, we have computed the 
diagonal Pad\'e approximants
\begin{equation}
P_{[M/M]}(C_{5,3})=\left[\,\,
\sum_{k=0}^{139} C_{5,3}^{(k)}\,\Big(\frac{\lambda}{\pi^2}\Big)^k\,\right]_{[M/M]}
\label{C53Pade}
\end{equation} 
for $M=26,40,68$ and compared them with the strong-coupling behavior of $C_{5,3}$
that can be obtained using the asymptotic form of $T_{3,3}$ and $T_{3,5}$ given in
(\ref{T35T33strong}), {\it{i.e.}}
\begin{equation}
C_{5,3}^{(\infty)}=\lim_{\lambda\to\infty} C_{5,3} \,\simeq\,\frac{T_{3,5}^{(\infty)}}{T_{3,3}^{(\infty)}}
=\frac{10N}{3}~.
\label{C53strong}
\end{equation}
The three Pad\'e approximants that we have computed and the strong-coupling result
(\ref{C53strong}) are plotted in Fig.~\ref{fig:8}, which shows that for increasing values of $M$ and for large values of $\lambda$ the numerical curves tend towards the expected
asymptotic value\,%
\footnote{The agreement could be improved by considering sub-leading contributions to the asymptotic value. These contributions can be obtained by keeping the sub-leading terms in the expansion of the matrix $\mathsf{X}$ for large $\lambda$.}.

\begin{figure}[ht]
\center{\includegraphics[scale=0.43]{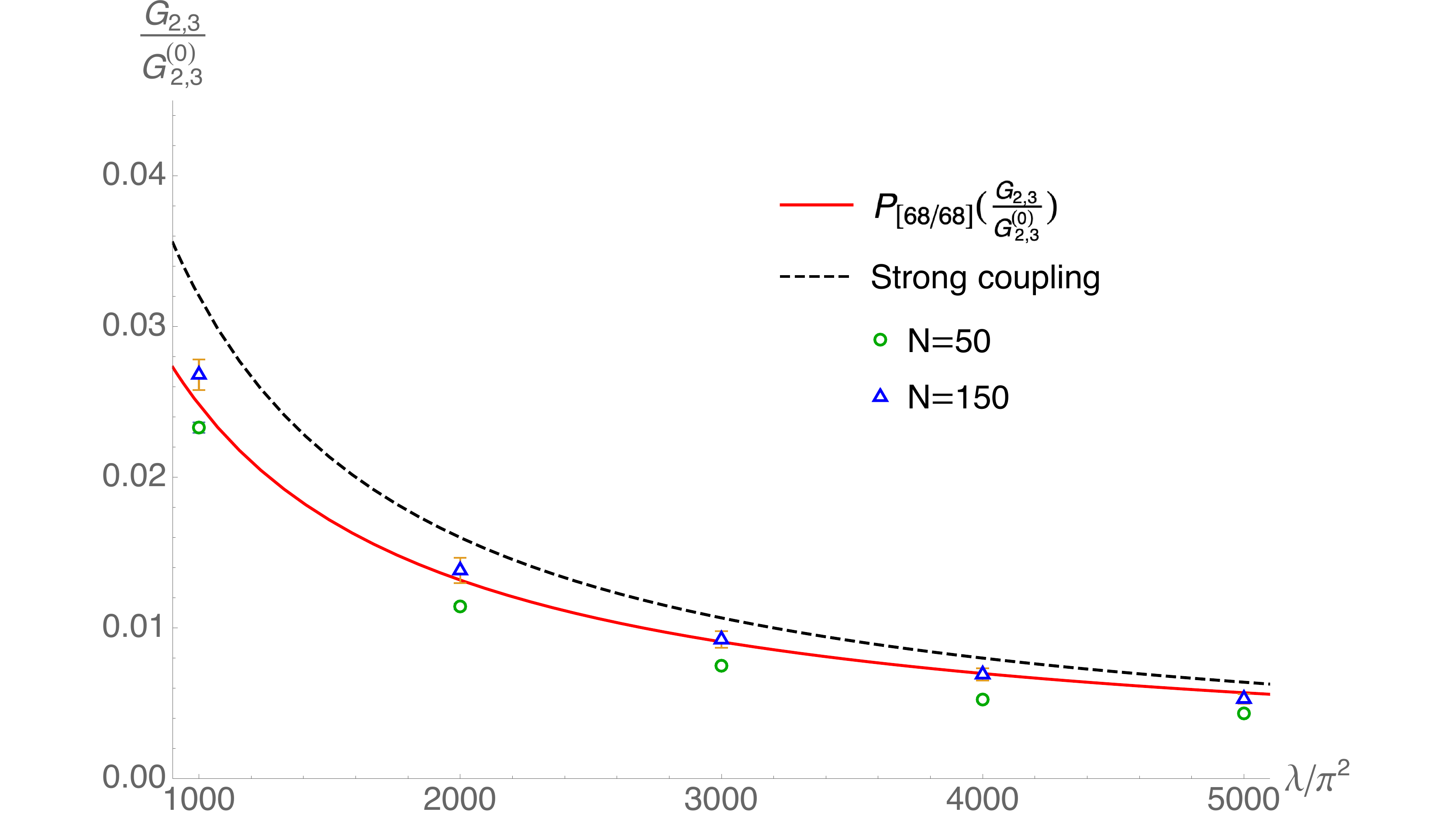}
\caption{Comparison between the Pad\'e curve at $M=68$ for the ratio 
$G_{2,3}/G_{2,3}^{(0)}$ (red curve), the large-$\lambda$ theoretical prediction (\ref{Delta23strong}) (black dashed curve) and the points from the Monte Carlo simulations at $N=50$ (green open circles) and at $N=150$ (blue open triangles).\label{fig:9}} 
    }  
\end{figure}

We have also computed the correlator $G_{2,3}$ in the $\mathbf{E}$ theory with a Monte Carlo simulation using the Metropolis-Hastings algorithm (see for instance \cite{brooks2011handbook}) for $N=50$ and $N=150$, along the same lines discussed
in \cite{Beccaria:2021hvt}. The results of these simulations are shown in Fig.~\ref{fig:9},
where we have also plotted the Pad\'e approximant for the ratio $G_{2,3}/G_{2,3}^{(0)}=(1+\Delta_{2,3})$ with $M=68$ and the large-$\lambda$ theoretical prediction (\ref{Delta23strong}).
We see that as $\lambda$ increases, the Monte Carlo points tend 
towards the Pad\'e curve which in turn tends towards the theoretical strong-coupling curve. As expected, the agreement of the Monte Carlo simulation is better for $N=150$ than for $N=50$.
We regard these numerical results as a strong evidence of the validity of our analysis.

\section{Three-point functions in the \texorpdfstring{$\mathbf{E}$}{} theory at large \texorpdfstring{$N$}{}: strong coupling results}
\label{secn:Estrong}
We now extend the results of the previous section by computing a generic 3-point function of single-trace operators 
of the $\mathbf{E}$ theory in the large-$N$ limit at strong coupling.

To do so, we first analyze in more detail the structure of the single-trace operators 
$O_n$ introduced in (\ref{newOn}). The mixing coefficients $C_{n,m}$, defined by 
imposing the  orthogonality relation (\ref{ortho1}), are given by
\begin{equation}
C_{n,m}=\frac{\big\langle\Omega_n \,O_m\big\rangle}{\big\langle\Omega_m \,O_m
\big\rangle }
\label{Cnm1} 
\end{equation}
where $n$ and $m$ are both even or both odd, with $m<n$. It is not difficult to realize 
that these coefficients in the end become rational homogeneous functions of the 
expectation values $T_{r,s}$ where $r$ and $s$ are both even if $n$ and $m$ are even, 
or both odd if $n$ and $m$ are odd.

As discussed in the previous sections, the form of these expectation values is explicitly 
known in the large-$N$ limit, both at weak and at strong coupling. Using this 
information we find that when $\lambda\to 0$, {\it{i.e.}} in the $\mathcal{N}=4$ SYM, 
the mixing coefficients are\,%
\footnote{This result follows directly from the findings of
\cite{Rodriguez-Gomez:2016cem} where it was shown that in the free Gaussian model at large $N$ the mixing coefficients of an operator of dimension $n$ are related to the coefficients of the $n$-th Chebyshev polynomial of the first kind. Using this information, the result in (\ref{C0nm}) immediately follows,
see also Section\,3.2 of \cite{Beccaria:2020hgy}.}
\begin{equation}
C_{n,m}^{(0)} = \lim_{\lambda\to0} C_{n,m}~\simeq~
\Big(\frac{N}{2}\Big)^{\frac{n-m}{2}}\,\binom{n}{\frac{n-m}{2}}~.
\label{C0nm}
\end{equation}
On the other hand when $\lambda\to \infty$ we can exploit the strong-coupling 
behavior of the expectation values $T_{r,s}$ given in (\ref{Tkleven}) and (\ref{Tklstrong}), 
and find that when the indices are even, the mixing coefficients remain unchanged at 
leading order, namely
\begin{equation}
C_{2k,2\ell}^{(\infty)}=\lim_{\lambda\to\infty} C_{2k,2\ell}\,\simeq\,C_{2k,2\ell}^{(0)}
\,\simeq \, \Big(\frac{N}{2}\Big)^{k-\ell}\,\binom{2k}{k-\ell}~,
\label{Cevenstrong}
\end{equation}
while when the indices are odd they acquire an extra simple numerical factor and become
\begin{equation}
C_{2k+1,2\ell+1}^{(\infty)}=\lim_{\lambda\to\infty} C_{2k+1,2\ell+1}\,\simeq\,
\frac{k+\ell+1}{2\ell+1}\,C_{2k+1,2\ell+1}^{(0)}
\,\simeq\,\frac{2k+1}{2\ell+1}\,\Big(\frac{N}{2}\Big)^{k-\ell}\,\binom{2k}{k-\ell}
~.
\label{Coddstrong}
\end{equation}
Note that for $k=2$ and $\ell=1$ we recover the explicit result in (\ref{C53strong}).
The relation (\ref{Coddstrong}), which we have checked in numerous examples even with very high values of $k$ and $\ell$, can be proven with a nested inductive method as shown in Appendix\,\ref{app:induction}\,%
\footnote{We warmly thank the anonymous referee for suggesting this proof.}.

To perform explicit calculations it is actually more convenient to express the operators $O_n$ in the basis of the vevless operators\,%
\footnote{Note that $\widehat{\Omega}_0=\widehat{\Omega}_1=0$ and $\widehat{\Omega}_{2k+1}=\Omega_{2k+1}$.}
\begin{equation}
\widehat{\Omega}_n=\Omega_n-\big\langle \Omega_n\big\rangle=\Omega_n-T_n~,
\label{Omegahat}
\end{equation}
 and write
\begin{equation}
O_n=\sum_{2\leq m\leq n} M_{n,m}\,\widehat{\Omega}_m
\label{OOmega}
\end{equation}
with $n$ and $m$ being both even or both odd. Comparing this with (\ref{newOn}), we easily see that the mixing matrices $M$ and $C$ are related as follows
\begin{equation}
M_{n,m}=\Big(\frac{1}{\mathbb{1}+C}\Big)_{n,m}~.
\label{MvsC}
\end{equation}
In the free theory at large $N$, using (\ref{C0nm}) one can show that
\begin{equation}
M_{n,m}^{(0)}=\lim_{\lambda\to 0}M_{n,m}\,\simeq\,\Big(\!\!-\frac{N}{2}\Big)^{\frac{n-m}{2}}\,\frac{n}{m}\,\binom{\frac{n+m-2}{2}}{\frac{n-m}{2}}
\label{Mnm0}
\end{equation}
and check that the expression in the right hand side is related to the coefficients of (suitably rescaled) Chebyshev polynomials, as originally pointed out in 
\cite{Rodriguez-Gomez:2016cem}. At strong coupling, instead, we have a different behavior depending on whether the indices are even or odd. In fact, from (\ref{Cevenstrong}), (\ref{Coddstrong}) and (\ref{MvsC}) we find, respectively,
\begin{equation}
M_{2k,2\ell}^{(\infty)}=\lim_{\lambda\to\infty}M_{2k,2\ell}\,\simeq\,M_{2k,2\ell}^{(0)}
\simeq \Big(\!\!-\frac{N}{2}\Big)^{k-\ell}\,\frac{k}{\ell}\,\binom{k+\ell-1}{k-\ell}
~,
\label{Mevenstrong}
\end{equation}
and
\begin{equation}
\begin{aligned}
M_{2k+1,2\ell+1}^{(\infty)}&=\lim_{\lambda\to\infty}M_{2k+1,2\ell+1}\,\simeq
\,\frac{2k}{k+\ell}\,M_{2k+1,2\ell+1}^{(0)}
\simeq\,\frac{2k+1}{2\ell+1}\, 
\Big(\!\!-\frac{N}{2}\Big)^{k-\ell}\,\frac{k}{\ell}\,\binom{k+\ell-1}{k-\ell}
~.
\end{aligned}
\label{Moddstrong}
\end{equation}

We are now in the position of computing the generic 3-point function 
of single-trace operators in 
the $\mathbf{E}$ theory. We start by considering $G_{2,2\ell+1}$. In this case we have to compute
\begin{equation}
\big\langle O_2\,O_{2\ell+1}\,O_{2p+1}\big\rangle 
\end{equation}
with $p=\ell+1$. Using (\ref{OOmega}) we immediately find
\begin{equation}
\big\langle O_2\,O_{2\ell+1}\,O_{2p+1}\big\rangle =
\big\langle \Omega_2\,O_{2\ell+1}\,O_{2p+1}\big\rangle =\sum_{r=1}^\ell\sum_{s=1}^p
M_{2\ell+1,2r+1}\,M_{2p+1,2s+1}\,T_{2,2r+1,2s+1}~,
\end{equation}
which, upon exploiting the relation (\ref{recursionT2}), can be rewritten as
\begin{align}
&\frac{1}{2}\left(N^2+1-2\lambda\,\partial_\lambda\mathcal{F}\right)
\sum_{r=1}^\ell\sum_{s=1}^p
M_{2\ell+1,2r+1}\,M_{2p+1,2s+1}\,T_{2r+1,2s+1}\notag\\
&~+\,\sum_{r=1}^\ell\sum_{s=1}^p
M_{2\ell+1,2r+1}\,M_{2p+1,2s+1}\,(r+s)\,T_{2r+1,2s+1}\label{3point2}\\
&~+\,\sum_{r=1}^\ell\sum_{s=1}^p
M_{2\ell+1,2r+1}\,M_{2p+1,2s+1}\,\lambda\,\partial_\lambda T_{2r+1,2s+1}~.\notag
\end{align}
The first line in the above expression vanishes because 
the double sum reconstructs the expectation value
$\big\langle O_{2\ell+1}\,O_{2p+1}\big\rangle$ which is zero due to the orthogonality condition
(\ref{ortho1}). 
Since we are interested in the strong coupling limit at large $N$, we can
replace the mixing coefficients $M$ and the expectation values $T$ with their asymptotic expressions $M^{(\infty)}$ and $T^{(\infty)}$ given, respectively, in (\ref{Moddstrong}) and (\ref{Tklstrong}). Having done this and exploiting the fact that $\lambda\,\partial_\lambda T^{(\infty)}_{2r+1,2s+1}=-T^{(\infty)}_{2r+1,2s+1}$, we see that also the third line of (\ref{3point2}) does not contribute at leading order when $\lambda\to\infty$ because it is proportional to $\big\langle O_{2\ell+1}\,O_{2p+1}\big\rangle$ which vanishes. Thus, we are left with
\begin{align}
\big\langle O_2\,O_{2\ell+1}\,O_{2p+1}\big\rangle &\simeq\,
\sum_{r=1}^\ell\sum_{s=1}^p
M^{(\infty)}_{2\ell+1,2r+1}\,M^{(\infty)}_{2p+1,2s+1}\,(r+s)\,T^{(\infty)}_{2r+1,2s+1}\notag\\
&=-\Big(\!\!-\frac{N}{2}\Big)^{\ell+p+1}\,\frac{16\pi^2}{\lambda}
\,\ell\,(2l+1)\,p\,(2p+1)\notag\\
&\quad\times\sum_{r=1}^{\ell}(-1)^r\frac{(\ell+r-1)!}{(\ell-r)!\,r!\,(r-1)!}~
\sum_{s=1}^{p}(-1)^s\frac{(p+s-1)!}{(p-s)!\,s!\,(s-1)!}~,
\label{O2lp}
\end{align}
where the second step follows from (\ref{Moddstrong}) and (\ref{Tklstrong}) and some simple algebraic manipulations. Using the identity
\begin{equation}
\sum_{r=1}^{\ell}(-1)^r\frac{(\ell+r-1)!}{(\ell-r)!\,r!\,(r-1)!}=(-1)^\ell~,
\label{identity}
\end{equation}
we finally obtain
\begin{equation}
\big\langle O_2\,O_{2\ell+1}\,O_{2p+1}\big\rangle\,\simeq\,
\Big(\frac{N}{2}\Big)^{\ell+p+1}\,\frac{16\pi^2}{\lambda}
\,\ell\,(2\ell+1)\,p\,(2p+1)~.
\label{3point3}
\end{equation}
Setting $p=\ell+1$ in this expression, we deduce that at strong coupling the 3-point function $G_{2,2\ell+1}$ is simply
\begin{equation}
\begin{aligned}
G_{2,2\ell+1}&\,\simeq\,\Big(\frac{N}{2}\Big)^{2\ell+2}\,\frac{16\pi^2}{\lambda}
\,\ell\,(\ell+1)(2\ell+1)\,(2\ell+3)\\[1mm]
&\,\simeq\,G_{2,2\ell+1}^{(0)} \,\frac{16\pi^2}{\lambda}
\,\ell\,(\ell+1)~.
\end{aligned}
\label{G2odd}
\end{equation}
Of course, for $\ell=1$ we retrieve the results of Section~\ref{secn:example} (see (\ref{O2O3O5quater})--(\ref{Delta23strong})).

Applying the same methods, one can show that
\begin{equation}
\big\langle O_{2m}\,O_{2\ell+1}\,O_{2p+1}\big\rangle\,\simeq\,
\Big(\frac{N}{2}\Big)^{m+\ell+p}\,\frac{16\pi^2}{\lambda}
\,m\,\ell\,(2\ell+1)\,p\,(2p+1)
\label{OOOgeneral}
\end{equation}
which is a simple generalization of (\ref{3point3}).
The details of the derivation of this result can be found in Appendix~\ref{app:3point}. If in (\ref{OOOgeneral}) we set
$m=k$ and $p=k+\ell$, we obtain
\begin{equation}
\begin{aligned}
G_{2k,2\ell+1}&\,\simeq\,\Big(\frac{N}{2}\Big)^{2k+2\ell}\,\frac{16\pi^2}{\lambda}
\,k\,\ell\,(k+\ell)\,(2\ell+1)\,(2k+2\ell+1)\\[1mm]
&\,\simeq\,G_{2k,2\ell+1}^{(0)} \,\frac{16\pi^2}{\lambda}
\,\ell\,(k+\ell)~,
\end{aligned}
\label{G2kodd}
\end{equation}
while if we set $m=k+\ell+1$ and $p=k$, we get
\begin{equation}
\begin{aligned}
G_{2k+1,2\ell+1}&\,\simeq\,\Big(\frac{N}{2}\Big)^{2k+2\ell+1}\,\frac{16\pi^2}{\lambda}
\,k\,\ell\,(2k+1)\,(2\ell+1)\,(k+\ell+1)\\[1mm]
&\,\simeq\,G_{2k+1,2\ell+1}^{(0)} \,\frac{16\pi^2}{\lambda}
\,k\,\ell~.
\end{aligned}
\label{G2oddodd}
\end{equation}
This concludes the analysis of the 3-point functions at strong coupling.

\section{Summary of results and conclusions}
\label{secn:concl}
The main result we have obtained in this paper is the strong coupling
behavior of the 3-point functions of single-trace operators $\mathcal{O}_n(x)$ of the $\mathbf{E}$ theory at large $N$. Writing
\begin{equation}
\big\langle \,\mathcal{O}_{n_1}(x_1)\,
\mathcal{O}_{n_2}(x_2)\,\overbar{\mathcal{O}}_{n_1+n_2}(y)\,\big\rangle
=\frac{G_{n_1,n_2}\phantom{\big|}}{
\big(4\pi^2(x_1-y)^2\big)^{n_1}\,\big(4\pi^2(x_2-y)^2\big)^{n_2}}~,
\label{3pointsinglebis}
\end{equation}
we have found for $\lambda\to\infty$ that
\begin{equation}
\begin{aligned}
G_{2k,2\ell}&\,\simeq\,G_{2k,2\ell}^{(0)}~,\\
G_{2k,2\ell+1}&\,\simeq\,G_{2k,2\ell+1}^{(0)}\,\frac{16\pi^2}{\lambda}
\,\ell\,(k+\ell)~,\\
G_{2k+1,2\ell+1}&\,\simeq\,G_{2k+1,2\ell+1}^{(0)}\,\frac{16\pi^2}{\lambda}
\,k\,\ell~,
\end{aligned}
\label{G3strong}
\end{equation}
where $G_{n_1,n_2}^{(0)}$ is the 3-point correlator in the $\mathcal{N}=4$ SYM given in
(\ref{Gn1n20}). On the other hand in \cite{Beccaria:2021hvt} it was shown that
at strong coupling and in the large-$N$ limit
the 2-point functions of the single-trace operators in the $\mathbf{E}$ theory are
\begin{equation}
\big\langle \,\mathcal{O}_{n}(x)\,\overbar{\mathcal{O}}_{n}(y)\,\big\rangle
=\frac{G_n\phantom{\big|}}{
\big(4\pi^2(x-y)^2\big)^{n}}~,
\label{2pointsinglebis}
\end{equation}
with
\begin{equation}
\begin{aligned}
G_{2k}&\,\simeq\,G_{2k}^{(0)}~,\\
G_{2k+1}&\,\simeq\,G_{2k+1}^{(0)}\,\frac{8\pi^2}{\lambda}
\,k\,(2k+1)~,
\end{aligned}
\label{G2strong}
\end{equation}
where $G_{n}^{(0)}$ is the 2-point correlator in the $\mathcal{N}=4$ SYM given in
(\ref{Gn0}). 
We can combine these results by defining the normalized operators
\begin{equation}
\widehat{\mathcal{O}}_{n}(x)=\frac{\mathcal{O}_n(x)}{\sqrt{G_n\phantom{\big|}}}
\end{equation}
whose correlators are 
\begin{equation}
\begin{aligned}
\big\langle \,\widehat{\mathcal{O}}_{n}(x)\,\widehat{\overbar{\mathcal{O}}}_{n}(y)\,\big\rangle
&=\frac{1\phantom{\big|}}{
\big(4\pi^2(x-y)^2\big)^{n}}\\
\big\langle \,\widehat{\mathcal{O}}_{n_1}(x_1)\,
\widehat{\mathcal{O}}_{n_2}(x_2)\,\widehat{\overbar{\mathcal{O}}}_{n_1+n_2}(y)\,\big\rangle
&=\frac{\widehat{G}_{n_1,n_2}\phantom{\big|}}{
\big(4\pi^2(x_1-y)^2\big)^{n_1}\,\big(4\pi^2(x_2-y)^2\big)^{n_2}}
\end{aligned}
\label{23pointsinglebis}
\end{equation}
with
\begin{equation}
\widehat{G}_{n_1,n_2}=\frac{G_{n_1,n_2}}{\sqrt{G_{n_1}\,G_{n_2}\,G_{n_1+n_2}\phantom{\big|}}}~.
\label{hatG}
\end{equation}
Using (\ref{G3strong}) and (\ref{G2strong}), it is easy to see that
at strong coupling
\begin{equation}
\begin{aligned}
\widehat{G}_{2k,2\ell}&\,\simeq\,\frac{1}{N}\,\sqrt{(2k)\,(2\ell)\,(2k+2\ell)}~,\\
\widehat{G}_{2k,2\ell+1}&\,\simeq\,\frac{1}{N}\,\sqrt{(2k)\,(2\ell)\,(2k+2\ell)}~,\\
\widehat{G}_{2k+1,2\ell+1}&\,\simeq\,\frac{1}{N}\,\sqrt{(2k)\,(2\ell)\,(2k+2\ell+2)}~.
\end{aligned}
\label{G3strongfinal}
\end{equation}
These normalized coefficients are part of the intrinsic data that characterize the conformal field theory under consideration in the strong-coupling regime. To our knowledge this is the first time that such a strong-coupling result is obtained in a 
$\mathcal{N}=2$ SYM theory.

We can rephrase our findings in a suggestive way by observing that the operators with even dimension $\widehat{\mathcal{O}}_{2k}(x)$ belong to the so-called ``untwisted'' sector, while those with odd dimension $\widehat{\mathcal{O}}_{2k+1}(x)$ are in the
``twisted'' sector. As explained in \cite{Beccaria:2021hvt,Billo:2021rdb} this 
terminology derives from the string construction of the $\mathbf{E}$ theory in terms
of a suitable orientifold projection of a two-node quiver model, which in turn can be engineered with a system of fractional D3-branes in a
$\mathbb{Z}_2$ orbifold of Type II B string theory \cite{Park:1998zh}. Indeed, by exploiting the open/closed string correspondence, one can prove that the even operators $\widehat{\mathcal{O}}_{2k}(x)$ correspond to open string configurations 
that are dual to closed string excitations of the untwisted sector which are 
even under the orbifold/orientifold parity, while the odd operators 
$\widehat{\mathcal{O}}_{2k+1}(x)$ are associated to open string configurations that are
dual to closed string modes of the $\mathbb{Z}_2$ twisted sector surviving 
the orbifold/orientifold projection \cite{Gukov:1998kk}. In computing the 3-point functions we have therefore two possibilities: a 3-point function with three untwisted operators or a 3-point function with one untwisted and two twisted operators. Calling, in an obvious notation,  $C_{U_1U_2U_3}$
and $C_{U_1T_2T_3}$ the coefficients appearing in the corresponding 3-point functions,
our strong-coupling results (\ref{G3strongfinal}) can be rewritten as 
\begin{subequations}
\begin{align}
C_{U_1U_2U_3}&\,\simeq\,\frac{1}{N}\,\sqrt{\phantom{\big|}\!d_{U_1}\,d_{U_2}\,d_{U_3}} ~,
\label{CUUU}\\[2mm]
 C_{U_1T_2T_3}&\,\simeq\,\frac{1}{N}\,\sqrt{\phantom{\big|}\!d_{U_1}\,(d_{T_2}-1)\,
(d_{T_3}-1)}~,
\label{CUTT}
\end{align}%
\end{subequations}
where $d$ denotes the conformal dimension of the operator. 

We point out that (\ref{CUUU}) is the same result found in $\mathcal{N}=4$ SYM. Indeed, the untwisted operators of the $\mathbf{E}$ theory behave in the same manner as the corresponding ones of the $\mathcal{N}=4$ SYM since they do not 
feel the $\mathbb{Z}_2$ orbifold/orientifold projection. By exploiting the AdS/CFT correspondence, the strong-coupling formula (\ref{CUUU}) has been explicitly confirmed long ago in \cite{Lee:1998bxa} with an explicit calculation of the 3-point functions
in $\mathrm{AdS}_5\times S^5$. On the other hand, (\ref{CUTT}) is a new strong-coupling result which would be very interesting to compare with a dual calculation in an AdS space with a $\mathbb{Z}_2$ orbifold/orientifold.

\vskip 1cm
\noindent {\large {\bf Note added}}
\vskip 0.2cm
While this paper was being reviewed, we have extended the calculation of the 3-point functions of scalar operators and of the corresponding structure constants to $\mathcal{N}=2$ quiver gauge theories with $M$ nodes, and showed that the strong-coupling behavior predicted by localization perfectly agrees with the one obtained with an holographic approach based on the AdS/CFT correspondence \cite{Billo:2022gmq,Billo:2022fnb}. The $\mathbf{E}$ theory considered in this paper is obtained from the two-node quiver theory by means of an orientifold projection.

\vskip 1cm
\noindent {\large {\bf Acknowledgments}}
\vskip 0.2cm
We would like to thank Francesco Galvagno, Marco Meineri and Igor Pesando 
for useful discussions.
This research is partially supported by the MUR PRIN contract 2020KR4KN2 ``String Theory as a bridge between Gauge Theories and Quantum Gravity'' and by
the INFN project ST\&FI
``String Theory \& Fundamental Interactions''. The work of A.P. is supported by INFN with a``Borsa di studio post-doctoral per fisici teorici".
\vskip 1cm
\begin{appendix}

\section{The mixing coefficients}
\label{app:mixing}

In this appendix we provide some details on mixing coefficients 
$\cC_{\mathbf{n}}^{\,\mathbf{m}}$ appearing in the recursive definition (\ref{Onis}) of the normal ordered operators $\cO_{\mathbf{n}}$, and derive a closed-form 
expression in terms of expectation values of non-normal ordered multi-trace operators. 

Note that the definition of the normal ordered operators $\cO_{\mathbf{n}}$ is such 
that they are orthogonal to all operators of lower dimension only. The 2-point 
functions $\vev{\cO_{\mathbf{n}} \cO_{\mathbf{m}}}$ with $|\mathbf{n}|=|\mathbf{m}|$ are instead not required to be diagonal. Of course, one could redefine the operators 
so as to orthogonalize them. However, like in the gauge theory where one does not
mix different trace structures with the same dimensions, also in the matrix model
we do not make this step and perform a Gram Schmidt procedure which is not complete. Thus, the formulas that we obtain in this way are not standard. Nevertheless we think that it may be useful to report them, also because they are valid in any matrix model and not only in the $\mathbf{E}$ theory considered in this paper.

The coefficients $\cC_{\mathbf{n}}^{\,\mathbf{m}}$ are determined by solving recursively the orthogonality conditions (\ref{OnOm}). Keeping $\mathbf{n}$ fixed, for all
$\mathbf{m}$'s such that $|\mathbf{m}|<|\mathbf{n}|$ we have to impose 
\begin{align}
	\label{ort}
		0 = \big\langle \cO_{\mathbf{m}} \,\cO_{\mathbf{n}} \big\rangle= \big\langle\cO_{\mathbf{m}} \,\Omega_{\mathbf{n}}\big\rangle -
		\sum_{|\mathbf{p}|<|\mathbf{n}|} \big\langle
		\cO_{\mathbf{m}}\, \cO_{\mathbf{p}}\big\rangle
		\,\cC_{\mathbf{n}}^{\,\mathbf{p}}~.     
\end{align}
If we have already determined the expression of operators with dimensions 
lower than $|\mathbf{n}|$ and their 2-point functions
\begin{align}
	\label{Gtwo}
		\cG_{\mathbf{m}\hspace{0.5pt}\mathbf{;}\hspace{0.5pt}\mathbf{p}} = \big\langle 
		\cO_{\mathbf{m}} \,\cO_{\mathbf{p}}\big\rangle~,
\end{align}
then we can solve the linear system for the unknowns $\cC_{\mathbf{n}}^{\,\mathbf{p}}$
given by the equations (\ref{ort}) for all values of $\mathbf{m}$ with $|\mathbf{m}| < |\mathbf{n}|$.	   

Due to the symmetry of the matrix model integral, the Gram-Schmidt procedure takes place separately in the sectors of even and odd operators. Here, for definiteness, we illustrate the formulas in the odd sector.
The lowest dimension odd operator is $\cO_3$, for which the expansion (\ref{Onis}) reduces simply to $\cO_3 = \Omega_3$,
so that
\begin{align}
	\label{cG3}
		\cG_{3\hspace{0.5pt}\mathbf{;}\hspace{0.5pt}3} = T_{3,3}~.
\end{align}
At dimension 5 there are two operators, $\cO_5$ and $\cO_{2,3}$, for which the expansion (\ref{Onis}) reads
\begin{align}
	\label{level5op}
		\cO_5 = \Omega_5 - \cC_5^{\,3}\, \cO_3~,~~~
		\cO_{2,3} = \Omega_{2,3} - \mathcal{C}_{2,3}^{\phantom{2,3}3}\, \cO_3~.
\end{align}
For each of these operators we have to impose a single orthogonality relation of the type (\ref{ort}), with $\mathbf{m} = 3$. This immediately determined the mixing coefficients which are
\begin{align}
	\label{level5}
		\cC_5^{\,3} = \frac{T_{3,5}}{T_{3,3}}~,~~~
		\mathcal{C}_{2,3}^{\phantom{2,3}3}= \frac{T_{3,2,3}}{T_{3,3}}~.
\end{align} 
With this information, we can now compute the 2-point correlators between the two operators of dimension 5, finding
 \begin{align}
	\label{G5}
		\cG_{5\hspace{0.5pt}\mathbf{;}\hspace{0.5pt}5} = \frac{1}{T_{3,3}} 
		\begin{vmatrix}
			T_{3,3} & T_{3,5} \cr
			T_{5,3} & T_{5,5}
		\end{vmatrix}~,~~~
		\cG_{5\hspace{0.5pt}\mathbf{;}\hspace{0.5pt}2,3} = \frac{1}{T_{3,3}} 
		\begin{vmatrix}
			T_{3,3} & T_{3,2,3} \cr
			T_{5,3} & T_{5,2,3}
		\end{vmatrix}~,~~~
		\cG_{2,3\hspace{0.5pt}\mathbf{;}\hspace{0.5pt}2,3} = \frac{1}{T_{3,3}} 
		\begin{vmatrix}
			T_{3,3} & T_{3,2,3} \cr
			T_{2,3,3} & T_{2,3,2,3}
		\end{vmatrix}~.
\end{align}

Proceeding recursively to operators of higher dimensions, we find that these formulas
can be generalized as follows. Let us introduce the matrices $\cT^{(k)}$ whose matrix elements are the 2-point functions of the operators $\Omega_{\mathbf{m}}$ with dimensions up to $k$, namely 
\begin{align}
	\label{Tk}
	\big[\cT^{(k)}\big]_{\mathbf{m}\hspace{0.5pt}\mathbf{;}\hspace{0.5pt}\mathbf{p}} = T_{\mathbf{p},\mathbf{m}}~~~~\text{with}~|\mathbf{m}|\,,\, |\mathbf{p}| \leq k~.	
\end{align}	
Then the mixing coefficient $\cC_{\mathbf{n}}^{\,\mathbf{m}}$ can be expressed in closed form as a ratio of determinants according to
\begin{align}
	\label{cCis}
		\cC_{\mathbf{n}}^{\,\mathbf{m}} = \frac{\left. \det \cT^{(m)}\right|_{\mathbf{m}\to \mathbf{n}} \phantom{\Big|}}{\det \cT^{(m)}\phantom{\Big|}}~,
\end{align}
where the notation in the numerator means that one has to replace the elements $T_{\mathbf{p},\mathbf{m}}$ in the column corresponding to the index $\mathbf{m}$ of $\cT^{(m)}$ with the quantities $T_{\mathbf{p},\mathbf{n}}$. For example, the mixing coefficient $\cC_7^{\,5}$ is given by
\begin{align}
	\label{C75}
		\cC_7^{\,5} =  \frac{\left. \det \cT^{(5)}\right|_{\mathbf{5}\to \mathbf{7}}  \phantom{\Big|}}{
		\det \cT^{(5)}}
		= 
		\begin{vmatrix}
			T_{3,3} & T_{3,7} & T_{3,2,3} \cr
			T_{5,3} & T_{5,7} & T_{5,2,3} \cr 
			T_{2,3,3} & T_{2,3,7} & T_{2,3,2,3}
		\end{vmatrix}
		\Bigg/
		\begin{vmatrix}
			T_{3,3} & T_{3,5} & T_{3,2,3} \cr
			T_{5,3} & T_{5,5} & T_{5,2,3} \cr 
			T_{2,3,3} & T_{2,3,5} & T_{2,3,2,3}
		\end{vmatrix}~.		
\end{align}
Finally, to describe the 2-point function $\cG_{\mathbf{n};\mathbf{n}^\prime}$ we introduce the matrix $\cT^{(\mathbf{n}\hspace{0.5pt}\mathbf{;}\hspace{0.5pt}
\mathbf{n}^\prime)}$ with elements
\begin{align}
	\label{Tnnp}
		\big[\cT^{(\mathbf{n};\mathbf{n}^\prime)}\big]_{\mathbf{p}\hspace{0.5pt}\mathbf{;}\hspace{0.5pt}\mathbf{q}} = T_{\mathbf{p},\mathbf{q}}
\end{align}
with $|\mathbf{p}|<|\mathbf{n}|$ or $\mathbf{p}=\mathbf{n}$ and 
$|\mathbf{q}|<|\mathbf{n}^\prime|$ or $\mathbf{q}=\mathbf{n}^\prime$. 
In practice, the matrix $\cT^{(\mathbf{n};\mathbf{n}^\prime)}$ is obtained from the matrix $\cT^{(|\mathbf{n}|-2)}$ introduced above, by adding one row with 
index $\mathbf{n}$ and one column of index $\mathbf{n}^\prime$. For instance,
\begin{align}
	\label{T523}
		\cT^{(5\hspace{0.5pt}\mathbf{;}\hspace{0.5pt}2,3)} = 
		\begin{pmatrix}
			~\cT^{(3)} & \vline & T_{3,2,3} \\
			\hline 
			\,T_{5,3} & \vline & T_{5,2,3}
		\end{pmatrix}
		=
		\begin{pmatrix}
			T_{3,3} & T_{3,2,3} \cr 
			T_{5,3} & T_{5,2,3}
		\end{pmatrix}~,
\end{align}
and
\begin{align}
	\label{T743}
		\cT^{(7\hspace{0.5pt}\mathbf{;}\hspace{0.5pt}4,3)} = 
		\begin{pmatrix}
			\cT^{(5)} & \vline & 
			\begin{matrix}
				T_{3,4,3}  \\
				T_{5,4,3}  \\
				T_{2,3,4,3}
			\end{matrix}
			\\
			\hline
			 \begin{matrix}
			 	T_{7,3} & T_{7,5} & T_{7,2,3}
			 \end{matrix} & \vline & T_{7,4,3}
		\end{pmatrix}=
		\begin{pmatrix}
		T_{3,3}&T_{3,5}&T_{2,2,3}&T_{3,4,3}\cr
		T_{5,3}&T_{5,5}&T_{5,2,3}&T_{5,4,3}\cr
		T_{2,3,3}&T_{2,3,5}&T_{2,3,2,3}&T_{2,3,4,3}\cr
		T_{7,3} & T_{7,5} & T_{7,2,3}&T_{7,4,3}
		\end{pmatrix}~.
\end{align}
Then, the 2-point functions are written as ratio of determinants as follows
\begin{align}
	\label{gratiodet}
		\cG_{\mathbf{n}\hspace{0.5pt}\mathbf{;}\hspace{0.5pt}\mathbf{n}^\prime} = \frac{\det \cT^{(\mathbf{n};\mathbf{n}^\prime)}\phantom{\Big|} }{\det \cT^{(|\mathbf{n}|-2)}\phantom{\Big|}}~.
\end{align}

\section{Recursion relations}
\label{app:recursion}
In this appendix we derive a relation valid at large $N$ 
between a correlator of the type 
\begin{align}
	\label{Tmlp}
		T_{2m,\mathbf{n}} = \big\langle \Omega_{2m,\mathbf{n}}\big\rangle
		= \big\langle \tr a^{2m}\, \tr a^{2\ell_1+1}\, \tr a^{2\ell_2+1}\ldots\big\rangle
\end{align}
where $\mathbf{n}=\{2\ell_1+1,2\ell_2+1,\ldots\}$ has an even number of odd components, and the correlator where the even insertion is missing, namely
\begin{align}
	\label{Tlp}
	T_{\mathbf{n}} = \big\langle\Omega_{\mathbf{n}}\big\rangle
	= \big\langle\tr a^{2\ell_1+1}\, \tr a^{2\ell_2+1}\ldots\big\rangle~.
\end{align}
The non-trivial relation that we find is instrumental in deriving the strong-coupling expression of the 3-point functions of one even and two odd operators 
which is reported in (\ref{OOOgeneral}) of the main text and is derived in 
Appendix~\ref{app:3point}. This relation can also be useful in evaluating the large-$N$ behavior of more general correlators in the matrix model.

\subsection*{Insertion of $\tr a^2$}
\label{subsec:m1case}
In the case $m=1$ it is quite straightforward to obtain an exact relation which is 
valid for any multi-trace $\Omega_\mathbf{n}$, not restricted to have only odd components. This is due to the special role played by the operator $\tr a^2$, which is the Gaussian weight of the matrix model.

Using the definition (\ref{vev}), we have 
\begin{align}
	\label{T2nis}
		T_{2,\mathbf{n}} = \big\langle \Omega_{2,\mathbf{n}}\big\rangle 
		= \frac{1}{\cZ} \int da\,\tr a^2 \, \Omega_{\mathbf{n}}~ 
		\rme^{-\tr a^2 - S_{\mathrm{int}}(a)}~.
\end{align}		      		
If we perform the rescaling 
\begin{align}
	\label{rescaM}
		a=\sqrt{\frac{8\pi^2 N}{\lambda}} M~,
\end{align}
then the quadratic term acquires a weight $-\frac{8\pi^2N}{\lambda}$ and the interaction 
action become independent of the coupling. More explicitly, we have
\begin{align}
	\label{numersc}
		\int da\,\tr a^2 \, \Omega_{\mathbf{n}}~ \rme^{-\tr a^2 - S_{\mathrm{int}}(a)} = 
		\Big(\frac{8\pi^2 N}{\lambda}\Big)^{\frac{N^2+1+|\mathbf{n}|}{2}} \!
		\int \!dM\,\tr M^2 \, \widetilde{\Omega}_{\mathbf{n}}~
		\rme^{-\frac{8\pi^2 N}{\lambda}\tr M^2 - \widetilde{S}_{\mathrm{int}}(M)}
\end{align}	
where $\widetilde{\Omega}_{\mathbf{n}} = \tr M^{n_1}\,\tr M^{n_2}\ldots$
and $\widetilde{S}_{\mathrm{int}}(M) = S_{\mathrm{int}}(a)$.
Similarly, the partition function becomes
\begin{align}
	\label{ZM}
		\cZ = \Big(\frac{8\pi^2 N}{\lambda}\Big)^{\frac{N^2- 1}{2}} \!
		\int \!dM\,
		\rme^{-\frac{8\pi^2 N}{\lambda}\tr M^2 - \widetilde{S}_{\mathrm{int}}(M)}~. 		
\end{align}
Thus, after the rescaling the correlator (\ref{T2nis}) is given by
\begin{align}
	\label{T2nisM}
		T_{2,\mathbf{n}} = \Big(\frac{8\pi^2 N}{\lambda}\Big)^{\frac{|\mathbf{n}|}{2} + 1}\,
		\frac{\displaystyle{\int\! dM\,\tr M^2 \, \widetilde{\Omega}_{\mathbf{n}}~ 
		\rme^{-\frac{8\pi^2 N}{\lambda}\tr M^2 - \widetilde{S}_{\mathrm{int}}(M)}}}{
\displaystyle{\int \!dM\,
		\rme^{-\frac{8\pi^2 N}{\lambda}\tr M^2 - \widetilde{S}_{\mathrm{int}}(M)}}}~. 
\end{align}	
Now we can trade the operator $\tr M^2$ appearing in the numerator for a derivative with respect to the coupling. With simple manipulations we find
\begin{align}
	\label{T2nisMbis}
		T_{2,\mathbf{n}} = \Big(\frac{8\pi^2 N}{\lambda}\Big)^{\frac{|\mathbf{n}|}{2}}\,
		\frac{\displaystyle{\lambda\,\partial_\lambda \!\int\! dM\,\widetilde{\Omega}_{\mathbf{n}}~ 
		\rme^{-\frac{8\pi^2 N}{\lambda}\tr M^2 - \widetilde{S}_{\mathrm{int}}(M)}}}{
\displaystyle{\int \!dM\,
		\rme^{-\frac{8\pi^2 N}{\lambda}\tr M^2 - \widetilde{S}_{\mathrm{int}}(M)}}}~. 
\end{align}
Now we change integration variables back to the original matrix $a$ getting
\begin{align}
	\label{T2nisMa}
	T_{2,\mathbf{n}} = 
	\Big(\frac{8\pi^2 N}{\lambda}\Big)^{\frac{|\mathbf{n}|}{2}}~
	\frac{\displaystyle{\lambda\,\partial_\lambda\bigg[\Big(\frac{\lambda}{8\pi^2N}\Big)^{\frac{N^2-1+|\mathbf{n}|}{2}}\int da\,\Omega_{\mathbf{n}}~ \rme^{-\tr a^2 - S_{\mathrm{int}}(a)} 
	\bigg]}}{\displaystyle{\Big(\frac{\lambda}{8\pi^2N}\Big)^{\frac{N^2-1}{2}}\,
	\mathcal{Z}}}~.
\end{align}
Taking the derivative, we remain with two contributions that can be recast as follows
\begin{align}
	\label{T2nis2p}
	T_{2,\mathbf{n}} = \frac{N^2-1+|\mathbf{n}|}{2} \,T_{\mathbf{n}} + \frac{1}{\cZ} 
	\,\lambda\,\partial_\lambda \big( \cZ\, T_{\mathbf{n}}\big)~,
 \end{align}
and further rearranged into 
\begin{align}
	\label{T2nisfin}
	T_{2,\mathbf{n}} = 
\frac{1}{2}\big(N^2-1+|\mathbf{n}|-2\lambda\,\partial_\lambda\mathcal{F}\big)\,T_{\mathbf{n}}
+\lambda\,\partial_\lambda T_{\mathbf{n}}	
\end{align}                                    
where $\cF = -\log \cZ$ is the free energy of the matrix model. For a 2-component vector $\mathbf{n}=\{n_1,n_2\}$, this formula becomes the one written in (\ref{recursionT2}) of the main text.

\subsection*{Insertion of $\tr a^{2m}$}
\label{subsec:genm}
When we consider the insertion of a generic even operator $\tr a^{2m}$, we can no longer use the strategy of the previous subsection because this operator does not appear in the free part of the matrix model action. In fact, in the $\mathbf{E}$ theory it does not appear at all in the action.

Using the recursion relations described in \cite{Billo:2017glv}, it is not difficult to see that
the following holds in the free theory ($\lambda=0$):
\begin{align}
	\label{t2mn}
		t_{2m,\mathbf{n}} \, \equiv\, 
		\big\langle \Omega_{2m,\mathbf{n}}\big\rangle_0 
		= t_{2m} \Big(1 + \frac{m(m+1)\, |\mathbf{n}|}{2N^2}+ O(1/N^4)\Big)\, t_{\mathbf{n}} ~,	
\end{align}
where $\mathbf{n}$ has an even number of odd entries. In the interacting case, we consider
\begin{align}
	\label{T2mn}
		T_{2m,\mathbf{n}} = \frac{\big\langle 
		\Omega_{2m,\mathbf{n}} \, 
		\rme^{-S_{\mathrm{int}}(a)}\big\rangle_0\phantom{\Big|}}
		{\big\langle
		\rme^{-S_{\mathrm{int}}(a)}\big\rangle_0\phantom{\Big|}}~.
\end{align}
Expanding in powers of the interaction action, up to quadratic order we have
\begin{align}
	\label{T2mnlin}
		T_{2m,\mathbf{n}} = t_{2m,\mathbf{n}} - T_{2m,\mathbf{n}}^{(1)} + T_{2m,\mathbf{n}}^{(2)}+\ldots
\end{align}
where
\begin{subequations}
\begin{align}
	\label{T12mn}
		T_{2m,\mathbf{n}}^{(1)} &= \big\langle 
		\Omega_{2m,\mathbf{n}}\, S_{\mathrm{int}}(a)\big\rangle_0 - 
		\big\langle\Omega_{2m,\mathbf{n}}\big\rangle_0 \,\big\langle
		S_{\mathrm{int}}(a)\big\rangle_0~,\\
T_{2m,\mathbf{n}}^{(2)}&=\frac{1}{2}\,\big\langle 
		\Omega_{2m,\mathbf{n}}\, S_{\mathrm{int}}^2(a)\big\rangle_0
		-\frac{1}{2}\,\big\langle 
		\Omega_{2m,\mathbf{n}}\big\rangle_0\, \big\langle S_{\mathrm{int}}^2(a)\big\rangle_0
		-T_{2m,\mathbf{n}}^{(1)} \,\big\langle S_{\mathrm{int}}^2(a)\big\rangle_0~.
\label{T2m2}
\end{align}
\end{subequations}
We take now into account the explicit form of the interaction action in the \textbf{E} theory, given in (\ref{Sint}), which only contains products of traces of two odd 
operators, and write as
\begin{align}
	\label{Sintf}
		S_{\mathrm{int}}(a) = \sum_{p=2}^\infty \sum_{q=1}^{p-1} 
		\Big(\frac{\lambda}{8\pi^2 N}\Big)^{p+1}
		 f_{p,q}\, \tr a^{2q+1} \tr a^{2(p-q)+1}~.
\end{align}
The coefficients $f_{p,q}$ can be determined by comparison with (\ref{Sint}), but their
expression is not relevant for our present computation. 
Inserting (\ref{Sintf}) into (\ref{T12mn}), we get
\begin{align}
	\label{T12mnis}
			T_{2m,\mathbf{n}}^{(1)} = \sum_{p=2}^\infty \sum_{q=1}^{p-1} \Big(\frac{\lambda}{8\pi^2 N}\Big)^{p+1}
			f_{p,q}\,
			\big(t_{2m,\mathbf{n},2q+1,2(p-q)+1} - t_{2m,\mathbf{n}}\, t_{2q+1,2(p-q)+1}\big)~. 
\end{align}
Now we can exploit the relation (\ref{t2mn}) obtaining
\begin{align}
	\label{T12mnis2}
		T_{2m,\mathbf{n}}^{(1)} & = t_{2m} \sum_{p=2}^\infty \sum_{q=1}^{p-1} \Big(\frac{\lambda}{8\pi^2 N}\Big)^{p+1} f_{p,q}\,
		\bigg\{\, t_{\mathbf{n},2q+1,2(p-q)+1} - t_{\mathbf{n}}\, t_{2q+1,2(p-q)+1}\\
		&~  + \frac{m(m+1)}{N^2} \Big[\Big(\frac{|\mathbf{n}|}{2} +p+1\Big)\, 
		t_{\mathbf{n},2q+1,2(p-q)+1} - \frac{|\mathbf{n}|}{2} \,  t_{\mathbf{n}}\, t_{2q+1,2(p-q)+1} \Big]
		+ O(1/N^4)\bigg\} ~. \notag
\end{align}
Rearranging the terms and noting that, inside the sum, $p+1$ can be traded for the action of $\lambda\,\partial_\lambda$, we see that the sum over $p$ and $q$ reduces to the one in the expansion (\ref{Sintf}) of $S_{\mathrm{int}}(a)$, and thus
we can write  
\begin{align}
	\label{T12mnl}
		T_{2m,\mathbf{n}}^{(1)} & = t_{2m} \bigg\{\Big[1 + \frac{m(m+1)}{N^2}
		\Big(\frac{|\mathbf{n}|}{2} + \lambda\,\partial_\lambda\Big)\Big]
		T_{\mathbf{n}}^{(1)}
		+ \frac{m(m+1)}{N^2}\, \lambda\,
		\partial_\lambda \big\langle S_{\mathrm{int}}(a)\big\rangle_0 \, t_{\mathbf{n}}
		+ O(1/N^4)\bigg\}~.
\end{align}
The same methods can be used to evaluate the terms quadratic in the interaction action given in (\ref{T2m2}), even if the algebra is a bit more involved. The result is
\begin{align}
T_{2m,\mathbf{n}}^{(2)} & = t_{2m} \bigg\{\Big[1 + \frac{m(m+1)}{N^2}
		\Big(\frac{|\mathbf{n}|}{2} + \lambda\,\partial_\lambda\Big)\Big]
		T_{\mathbf{n}}^{(2)}\label{T2m2bis}\\
		&~\,+\frac{m(m+1)}{N^2}\Big[\frac{1}{2}\,\lambda\,
		\partial_\lambda \big\langle S_{\mathrm{int}}^2(a)\big\rangle_0 \, t_{\mathbf{n}}
		-\frac{1}{2}\,\lambda\,
		\partial_\lambda \big\langle S_{\mathrm{int}}(a)\big\rangle_0^2 \, t_{\mathbf{n}}+\lambda\,
		\partial_\lambda \big\langle S_{\mathrm{int}}(a)\big\rangle_0 \, T_{\mathbf{n}}^{(1)}\Big]
		+O(1/N^4)\bigg\}~.		
		\notag
\end{align}
Putting everything together, up to terms that are sub-leading at large $N$ we obtain
\begin{align}
T_{2m,\mathbf{n}}& =t_{2m} \Big[1 + \frac{m(m+1)}{N^2}
		\Big(\frac{|\mathbf{n}|}{2} + \lambda\,\partial_\lambda\Big)\Big]
		\big(t_{\mathbf{n}}-T_{\mathbf{n}}^{(1)}+T_{\mathbf{n}}^{(2)}\big)\label{T2mter}\\
		&~-t_{2m}\,\frac{m(m+1)}{N^2}\Big[
		\lambda\,
		\partial_\lambda \big\langle S_{\mathrm{int}}(a)\big\rangle_0 \, 
		\big(t_{\mathbf{n}}-T_{\mathbf{n}}^{(1)}\big)
		-\frac{1}{2}\,\lambda\,
		\partial_\lambda \big\langle S_{\mathrm{int}}^2(a)\big\rangle_0 \,t_{\mathbf{n}}
		+\frac{1}{2}\,\lambda\,
		\partial_\lambda \big\langle S_{\mathrm{int}}(a)\big\rangle_0^2 \, t_{\mathbf{n}}\Big]
		\notag\\
		&~+\ldots\notag~.
\end{align}
The expression in square brackets in the second line above can be rewritten in terms of
the logarithmic derivative of the free energy, which is
\begin{align}
\lambda\,\partial_\lambda\mathcal{F}&=-\frac{\lambda\,\partial_\lambda \big\langle
		\rme^{-S_{\mathrm{int}}(a)}\big\rangle_0\phantom{\Big|}}{\big\langle
		\rme^{-S_{\mathrm{int}}(a)}\big\rangle_0\phantom{\Big|}} \label{freeenergy}\\
&=\lambda\,\partial_\lambda\big\langle S_{\mathrm{int}}(a)\big\rangle_0
-\frac{1}{2}\,\lambda\,
		\partial_\lambda \big\langle S_{\mathrm{int}}^2(a)\big\rangle_0
		+\frac{1}{2}\,\lambda\,
		\partial_\lambda \big\langle S_{\mathrm{int}}(a)\big\rangle_0^2+\ldots~.
		\notag
\end{align}
Indeed, by expanding
\begin{equation}
-t_{2m}\,\frac{m(m+1)}{N^2}\,
\lambda\,\partial_\lambda\mathcal{F}\,T_{\mathbf{n}}
\end{equation}
up to the second order in the interaction action, we precisely obtain the
second line of (\ref{T2mter}). This result clearly suggests its completion to all orders.
Altogether we get
\begin{align}
	\label{T12mntot}
		T_{2m,\mathbf{n}} & = t_{2m} \Big[1 + \frac{m(m+1)}{N^2}\Big(
		\frac{|\mathbf{n}|}{2} 
		- \lambda\,\partial_\lambda \cF + \lambda\,\partial_\lambda\Big)+ O(1/N^4)
		\Big] T_{\mathbf{n}}~.
\end{align}
In the particular case $\mathbf{n}=0$, this reduces to the following expression for the expectation value of an even trace:
\begin{align}
	\label{T2mis}
		T_{2m} = t_{2m} \Big(1 - \frac{m(m+1)}{N^2}\, \lambda\,\partial_\lambda 
		\cF + O(1/N^4)\Big)~,
\end{align} 
which was already given in Eq.\,(3.47) of \cite{Beccaria:2021hvt}. 
Using (\ref{T2mis}) inside (\ref{T12mntot}), we can rewrite the latter as 
\begin{align}
	\label{T12mntot1}
	T_{2m,\mathbf{n}} = T_{2m} \Big[1 + \frac{m(m+1)}{N^2}\Big(
	\frac{|\mathbf{n}|}{2} 
	+ \lambda\,\partial_\lambda\Big)+ O(1/N^4)\Big] T_{\mathbf{n}}~.
\end{align}
If in (\ref{T12mntot}) we take $m=1$ we retrieve, up to corrections of order $1/N^4$, the exact formula given above in (\ref{T2nisfin}).

\section{Proof of \texorpdfstring{Eq.\,(\ref{Coddstrong})}{}}
\label{app:induction}

Eq.\,(\ref{Coddstrong}) can be proved with a (nested) induction argument, namely we fix $k$ and, assuming it is valid for any $k'<k$, we prove that is valid also for $k$.

First of all, under our hypothesis we show that at strong coupling the following relation holds
\begin{equation}
\big\langle \Omega_{2k+1}\,O_{2\ell+1}\big\rangle\simeq
\frac{4\pi^2}{\lambda}\,\Big(\frac{N}{2}\Big)^{k+\ell+1}\,\frac{2\ell\,(2\ell+1)\,(2k+1)!}{(k-\ell)!\,(k+\ell)!}~.
\label{C1}
\end{equation}
We can prove this by induction on $\ell$. We have
\begin{align}
\big\langle \Omega_{2k+1}\,O_{2\ell+1}\big\rangle&=
\big\langle \Omega_{2k+1}\,\Omega_{2\ell+1}\big\rangle-\sum_{m<\ell}C_{2\ell+1,2m+1}\,
\big\langle \Omega_{2k+1}\,O_{2m+1}\big\rangle\notag\\
&\simeq\frac{4\pi^2}{\lambda}\,\Big(\frac{N}{2}\Big)^{k+\ell+1}\,\frac{(2k+1)!}{k!\,(k-1)!}
\,\frac{(2\ell+1)!}{\ell!\,(\ell-1)!}\,\frac{1}{k+\ell}\label{C2}\\[2mm]
&\quad-\sum_{m<\ell}\bigg[\frac{2\ell+1}{2m+1}\,\Big(\frac{N}{2}\Big)^{\ell-m}\binom{2\ell}{\ell-m}\bigg]\,\bigg[\frac{4\pi^2}{\lambda}\,\Big(\frac{N}{2}\Big)^{k+m+1}\,\frac{2m\,(2m+1)\,(2k+1)!}{(k-m)!\,(k+m)!}\bigg]~.\notag
\end{align}
Here the second line follows from (\ref{Tklstrong}), while the two square brackets in the last line
arise from using, respectively, (\ref{Coddstrong}) and (\ref{C1}) under our nested induction hypothesis.
With some straightforward algebra, we can recast (\ref{C2}) in the following form
\begin{align}
\big\langle \Omega_{2k+1}\,O_{2\ell+1}\big\rangle\simeq
\frac{4\pi^2}{\lambda}\,\Big(\frac{N}{2}\Big)^{k+\ell+1}(2k+1)\,(2\ell+1)
\bigg[\frac{k\,\ell}{k+\ell}\binom{2k}{k}\binom{2\ell}{\ell}
\!-\!\sum_{m<\ell}2m\binom{2k}{k-m}\binom{2\ell}{\ell-m}\bigg]~.
\label{C3}
\end{align}
Using the following binomial identity
\begin{equation}
\sum_{m\leq\ell}2m\binom{2k}{k-m}\binom{2\ell}{\ell-m}=\frac{k\,\ell}{k+\ell}\binom{2k}{k}\binom{2\ell}{\ell}~,
\label{C4}
\end{equation}
we see that most of the terms cancel and we remain with
\begin{align}
\big\langle \Omega_{2k+1}\,O_{2\ell+1}\big\rangle&\simeq
\frac{4\pi^2}{\lambda}\,\Big(\frac{N}{2}\Big)^{k+\ell+1}(2k+1)\,(2\ell+1)
\,2\ell\,\binom{2k}{k-\ell}\notag\\
&\simeq\frac{4\pi^2}{\lambda}\,\Big(\frac{N}{2}\Big)^{k+\ell+1}\,\frac{2\ell\,(2\ell+1)\,(2k+1)!}{(k-\ell)!\,(k+\ell)!}
\label{C5}
\end{align}
which is (\ref{C1}). Then, using this result it is easy to see that
\begin{align}
C_{2k+1,2\ell+1}^{(\infty)}&=\lim_{\lambda\to\infty}
\frac{\big\langle \Omega_{2k+1}\,O_{2\ell+1}\big\rangle}{\big\langle \Omega_{2\ell+1}
\,O_{2\ell+1}\big\rangle}\simeq
\frac{\displaystyle{\frac{4\pi^2}{\lambda}\,\Big(\frac{N}{2}\Big)^{k+\ell+1}\,\frac{2\ell\,(2\ell+1)\,(2k+1)!}{(k-\ell)!\,(k+\ell)!}}}{\displaystyle{\frac{4\pi^2}{\lambda}\,\Big(\frac{N}{2}\Big)^{2\ell+1}\,\frac{2\ell\,(2\ell+1)\,(2\ell+1)!}{(2\ell)!}}}\notag\\
&\simeq\frac{2k+1}{2\ell+1}\,\Big(\frac{N}{2}\Big)^{k-\ell}\,\binom{2k}{k-\ell}~,
\label{C6}
\end{align}
which is the relation appearing in (\ref{Coddstrong}) of the main text.

\section{The calculation of \texorpdfstring{$\big\langle O_{2m}\,O_{2\ell+1}\,O_{2p+1}\big\rangle$}{} }
\label{app:3point}
Here we provide some details on the computation of the 3-point function 
\begin{equation}
\big\langle O_{2m}\,O_{2\ell+1}\,O_{2p+1} \big\rangle
\label{Omlp}
\end{equation}
in the strong-coupling limit at large $N$ for the $\mathbf{E}$ theory.

Using (\ref{OOmega}) we have
\begin{align}
\big\langle O_{2m}\,O_{2\ell+1}\,O_{2p+1} \big\rangle 
= \sum_{n=1}^{m}\sum_{r=1}^{\ell}\sum_{s=1}^p M_{2m,2n}
\,M_{2\ell+1,2r+1}\,M_{2p+1,2s+1}
\big(T_{2n,2r+1,2s+1}-T_{2n}T_{2r+1,2s+1}\big)
~.
\label{Omlp1}
\end{align}
Exploiting the relation (\ref{T12mntot1}) for the case at hand, we easily see that
\begin{equation}
T_{2n,2r+1,2s+1}-T_{2n}T_{2r+1,2s+1}\,\simeq\,
\frac{n(n+1)}{N^2}\,T_{2n}\,\Big[(r+s+1)\,T_{2r+1,2s+1}
+\lambda\,\partial_\lambda T_{2r+1,2s+1}\Big]~.
\end{equation}
Inserting this result into (\ref{Omlp1}), we realize that the factor $(r+s+1)$ can be replaced by $(r+s)$ since the 1 gives a vanishing contribution. Indeed, the resulting
double sum over $r$ and $s$ factorizes and reproduces the expectation value 
$\big\langle O_{2\ell+1}\,O_{2p+1} \big\rangle$ which is zero due to the 
orthogonality condition (\ref{OnOm}). Thus, the correlator (\ref{Omlp1}) becomes
\begin{align}
\big\langle O_{2m}\,O_{2\ell+1}\,O_{2p+1} \big\rangle 
&\,\simeq\frac{1}{N^2}\sum_{n=1}^m \sum_{r=1}^\ell\sum_{s=1}^p
n(n+1) \,T_{2n} M_{2m,2n}
 M_{2\ell+1,2r+1} M_{2p+1,2s+1}\,(r+s)\,T_{2r+1,2s+1}
\notag\\
&\quad+\frac{1}{N^2}\sum_{n=1}^m\sum_{r=1}^\ell\sum_{s=1}^pn(n+1) \,T_{2n} 
M_{2m,2n}
 M_{2\ell+1,2r+1}M_{2p+1,2s+1}\,\lambda\,\partial_\lambda T_{2r+1,2s+1}
~.
\label{eq:expansion}
\end{align}
To find the strong-coupling limit at large $N$ we can replace the mixing coefficients 
$M$ and the expectation values $T$ in the above formula with the corresponding asymptotic 
expressions $M^{(\infty)}$ and $T^{(\infty)}$ for $\lambda\to\infty$. Exploiting the fact that $\lambda\,\partial_{\lambda}T_{2r+1,2s+1}^{(\infty)}=-T_{2r+1,2s+1}^{(\infty)}$, we observe that the second line of (\ref{eq:expansion}) 
does not contribute in this limit since it is proportional to $\langle O_{2\ell+1}O_{2p+1} \rangle$ which vanishes. We are then left with
\begin{align}
\langle O_{2m}O_{2\ell+1}O_{2p+1} \rangle \simeq 
\frac{1}{N^2}\sum_{n=1}^m \sum_{r=1}^\ell\sum_{s=1}^p
n(n+1) \,T_{2n}^{(\infty)} M_{2m,2n}^{(\infty)}
 M_{2\ell+1,2r+1}^{(\infty)} M_{2p+1,2s+1}^{(\infty)}\,(r+s)\,T_{2r+1,2s+1}^{(\infty)}~.
\end{align}
Notice that the sum over $n$ and the sums over $r$ and $s$ factorize. In particular the double sum over $r$ and $s$ is precisely what appears in the first line of (\ref{O2lp}). Therefore, following
the same steps described in Section~\ref{secn:Estrong} that lead to (\ref{3point3}), we have
\begin{align}
\sum_{r=1}^\ell\sum_{s=1}^p
 M_{2\ell+1,2r+1}^{(\infty)} M_{2p+1,2s+1}^{(\infty)}\,(r+s)\,T_{2r+1,2s+1}^{(\infty)}
 \, \simeq\, \Big(\frac{N}{2}\Big)^{\ell+p+1}\,\frac{16\pi^2}{\lambda}\,\ell\,(2\ell+1)\,p\,(2p+1)~.
\label{sumrs}
\end{align}
Let's now consider the remaining sum over $n$. To evaluate it, we first recall that
\begin{equation}
T_{2n}^{(\infty)}\,\simeq\,\frac{N^{n+1}}{2^n}\,\frac{(2n)!}{n!\,(n+1)!}
\end{equation}
(see Eq.\,(3.3) of \cite{Beccaria:2020hgy}); then using (\ref{Mevenstrong}) we get
\begin{align}
\frac{1}{N^2}\sum_{n=1}^m
n(n+1)\, T_{2n}^{(\infty)} M_{2m,2n}^{(\infty)}\,\simeq\,
m\,\Big(\frac{N}{2}\Big)^{m-1}\sum_{n=1}^{m}\frac{(-1)^{m+n}(m+n-1)!}{(m-n)!\,n!\,(n-1)!} = m\,
\Big(\frac{N}{2}\Big)^{m-1}
\label{sumn}
\end{align}
where in the last step we have exploited the identity (\ref{identity}). 
Finally, multiplying (\ref{sumrs}) and (\ref{sumn}) we obtain
\begin{equation}
\big\langle O_{2m}\,O_{2\ell+1}\,O_{2p+1}\big\rangle\,\simeq\,
\Big(\frac{N}{2}\Big)^{m+\ell+p}\,\frac{16\pi^2}{\lambda}
\,m\,\ell\,(2\ell+1)\,p\,(2p+1)
\label{OOOfinal}
\end{equation}
which is Eq.\,(\ref{OOOgeneral}) of the main text.

\section{Sub-leading corrections}
\label{app:subleading}
In this appendix we study the sub-leading corrections in the large-$N$ expansion of 
the correlators of single-trace operators with odd dimension. 
Even if this material is not directly relevant
for the main purpose of this paper, it may be useful in future developments.

As a first step, we briefly recall that in the free theory, {\it{i.e.}} in $\mathcal{N}=4$ SYM, 
at the leading order (LO) in the large-$N$ expansion one has (see \cite{Beccaria:2020hgy} for details)
\begin{equation}
\big\langle\Omega_{2\ell_1+1}\,\Omega_{2\ell_2+1}\big\rangle_0
\,\simeq\,H^{\mathrm{LO}}_{\ell_1,\ell_2}\,\beta_{\ell_1}\,\beta_{\ell_2}
\label{2pointL}
\end{equation}
where
\begin{align}
\beta_{\ell} = \frac{N^{\ell+1/2}}{\sqrt{2}}\,\frac{\ell\,(2\ell+1)!!}{(\ell+1)!} \quad\mbox{and}\quad
H^{\mathrm{LO}}_{\ell_1,\ell_2} = \frac{1}{1+\ell_1+\ell_2} ~.
\label{betaH0}
\end{align}
More generally, at large $N$ the correlator involving an even number $n$ of odd operators takes the form
\begin{align}
\big\langle\Omega_{2\ell_1+1}\,\Omega_{2\ell_2+1}\cdots\Omega_{2\ell_n+1}\big\rangle_0
\,\simeq\, \mathcal{H}^{\mathrm{LO}}_{\ell_1,\ell_2,\ldots,\ell_n}\,\prod_{i=1}^{n}\beta_{\ell_i}
\label{tn}
\end{align}
where $\mathcal{H}^{\mathrm{LO}}_{\ell_1,\ell_2,\cdots,\ell_n}$ 
represents the total Wick contraction computed with the ``propagator'' 
$H^{\mathrm{LO}}_{\ell_i,\ell_j}$. For example, if $n=4$ we have
\begin{align}
\label{H04}
\mathcal{H}^{\mathrm{LO}}_{\ell_1,\ell_2,\ell_3,\ell_4} = H^{\mathrm{LO}}_{\ell_1,\ell_2}\,H^{\mathrm{LO}}_{\ell_3,\ell_4} + H^{\mathrm{LO}}_{\ell_1,\ell_3}\,
H^{\mathrm{LO}}_{\ell_2,\ell_4}+ H^{\mathrm{LO}}_{\ell_1,\ell_4}\,
H^{\mathrm{LO}}_{\ell_2,\ell_3}~.
\end{align}

Let us now consider the next-to-leading (NLO) corrections in the large-$N$ expansion. 
Going to order $1/N^2$, the 2-point correlator (\ref{2pointL}) becomes
\begin{align}
\big\langle\Omega_{2\ell_1+1}\,\Omega_{2\ell_2+1}\big\rangle_0 \,\simeq\, 
H^{\mathrm{LO}}_{\ell_1,\ell_2}\,\beta_{\ell_1}\,\beta_{\ell_2}+H^{\mathrm{NLO}}_{\ell_1,\ell_2}\,\beta_{\ell_1}\,\beta_{\ell_2}
\label{2pointfree}
\end{align}
where 
\begin{align}
H^{\mathrm{NLO}}_{\ell_1,\ell_2} = \frac{1}{12N^2}\Big[\,\sum_{i=1}^{2}(\ell_i-1)(\ell_i-3)+(\ell_1-1)(\ell_2-1)-20\,\Big]~. \label{H1}
\end{align}
This term can be interpreted as the $O\big(1/N^2\big)$ correction to the propagator 
$H^{\mathrm{LO}}_{\ell_1,\ell_2}$. Therefore, we can promote 
$\mathcal{H}^{\mathrm{LO}}_{\ell_1,\ell_2,\cdots,\ell_n}$ to
$\mathcal{H}^{\mathrm{LO+NLO}}_{\ell_1,\ell_2,\cdots,\ell_n}$
by performing the Wick contractions with the propagator $H^{\mathrm{LO}}_{\ell_i,\ell_j}+H^{\mathrm{NLO}}_{\ell_i,\ell_j}$. However, if we compute the 4-point correlator
$\big\langle \Omega_{2\ell_1+1}\,\Omega_{2\ell_2+1}\,\Omega_{2\ell_3+1}\,\Omega_{2\ell_4+1}
\big\rangle_0$, we see that its $O\big(1/N^2\big)$ correction is not entirely captured by
$\mathcal{H}^{\mathrm{LO+NLO}}_{\ell_1,\ell_2,\ell_3,\ell_4}$ and one has to include an extra NLO term given by
\begin{align}
\label{V4}
V^{\mathrm{NLO}}_{\ell_1,\ell_2,\ell_3,\ell_4} = \frac{1}{N^2}\Big(\sum_{i=1}^{4}\ell_i+4\Big)~. 
\end{align}
Indeed, one can check that 
\begin{align}
\label{4pointfree}
\big\langle \Omega_{2\ell_1+1}\,\Omega_{2\ell_2+1}\,\Omega_{2\ell_3+1}\,\Omega_{2\ell_4+1}
\big\rangle_0\,\simeq\,
\Big(\mathcal{H}^{\mathrm{LO+NLO}}_{\ell_1,\ell_2,\ell_3,\ell_4}+V^{\mathrm{NLO}}_{\ell_1,\ell_2,\ell_3,\ell_4}
\Big)\prod_{i=1}^4\beta_{\ell_i}~.
\end{align}
We have verified that, remarkably, this quartic ``vertex'' together with the corrected propagator is enough to generate the NLO terms in all higher correlators using Wick's rule and that no higher vertices are needed at this order. 
For example the correlator of six operators up to NLO reads
\begin{align}
\big\langle \Omega_{2\ell_1+1}\,\Omega_{2\ell_2+1}\,&\Omega_{2\ell_3+1}\,\Omega_{2\ell_4+1}
\,\Omega_{2\ell_5+1}\,\Omega_{2\ell_6+1}\big\rangle_0\,\simeq\,
\Big(\mathcal{H}^{\mathrm{LO+NLO}}_{\ell_1,\ell_2,\ell_3,\ell_4,\ell_5,\ell_6}+
H^{\mathrm{LO}}_{\ell_1,\ell_2}\,V^{\mathrm{NLO}}_{\ell_3,\ell_4,\ell_5,\ell_6} \notag\\
&+
H^{\mathrm{LO}}_{\ell_1,\ell_3}\,V^{\mathrm{NLO}}_{\ell_2,\ell_4,\ell_5,\ell_6} 
+\,\ldots\,+
H^{\mathrm{LO}}_{\ell_5,\ell_6}\,V^{\mathrm{NLO}}_{\ell_1,\ell_2,\ell_3,\ell_4} \Big)
\prod_{i=1}^{6}\beta_{\ell_i}~.
\label{6point}
\end{align}

This structure becomes more transparent if we use the basis of the normalized normal-ordered operators of the free theory defined as
\begin{equation}
\omega_{2\ell+1}=\frac{1}{\sqrt{G^{(0)}_{2\ell+1}}}\sum_{r=1}^{\ell}
M^{(0)}_{2\ell+1,2r+1}\,\Omega_{2r+1}
\label{omega}
\end{equation}
where $M^{(0)}_{2\ell+1,2r+1}$ and $G^{(0)}_{2\ell+1}$ are given, respectively, 
in (\ref{Mnm0}) and (\ref{Gn0}). By construction, these operators are 
orthogonal to each other at LO, but if we include the NLO corrections their 2-point correlator acquires a non-diagonal term and becomes
\begin{align}
\label{2free}
\big\langle \omega_{2\ell_1+1}\,\omega_{2\ell_2+1}\big\rangle_{0} 
\,\simeq\, \delta_{\ell_1,\ell_2}+\frac{1}{N^2}\,f^{\mathrm{NLO}}_{\ell_1,\ell_2}
\end{align}
with
\begin{align}
\label{f}
f^{\mathrm{NLO}}_{\ell_1,\ell_2} = \frac{\sqrt{(2\ell_1+1)(2\ell_2+1)}}{24}\Big(\sum_{i=1}^{2}\ell_i(\ell_i+1)-14\Big)\,\Big(\sum_{i=1}^{2}\ell_i(\ell_i+1)\Big)~.
\end{align}
Using the $\omega$-basis, one finds that the quartic vertex (\ref{V4}) takes the following form 
\begin{align}
\label{Rvertex}
v^{\mathrm{NLO}}_{\ell_1,\ell_2,\ell_3,\ell_4} = \Big(\prod_{i=1}^4\sqrt{2\ell_i+1}\Big)\,\Big(
\sum_{i=1}^4\ell_i(\ell_i+1)\Big)~.
\end{align}

Let's now analyze the sub-leading corrections in the $\mathbf{E}$ theory. 
As proved in \cite{Beccaria:2020hgy}, at LO the 2-point correlators of the operators
(\ref{omega}) are
given by
\begin{align}
\label{2pointINT}
\big\langle \omega_{2\ell_1+1}\,\omega_{2\ell_2+1}\big \rangle \,\simeq\,
\Big(\frac{1}{\mathbb{1}-\Xx}\Big)_{\ell_1,\ell_2} \equiv\, \mathsf{D}^{\mathrm{LO}}
_{\ell_1,\ell_2}
\end{align} 
where $\Xx$ is an infinite matrix whose elements are defined in (\ref{Xkl}). Higher point correlators of $\omega$ operators are described at LO by Feynman diagrams constructed only with this ``propagator'' and no interaction vertices. 

\begin{figure}[t]
\vspace{-2cm}
\center{
\includegraphics[scale=0.40]{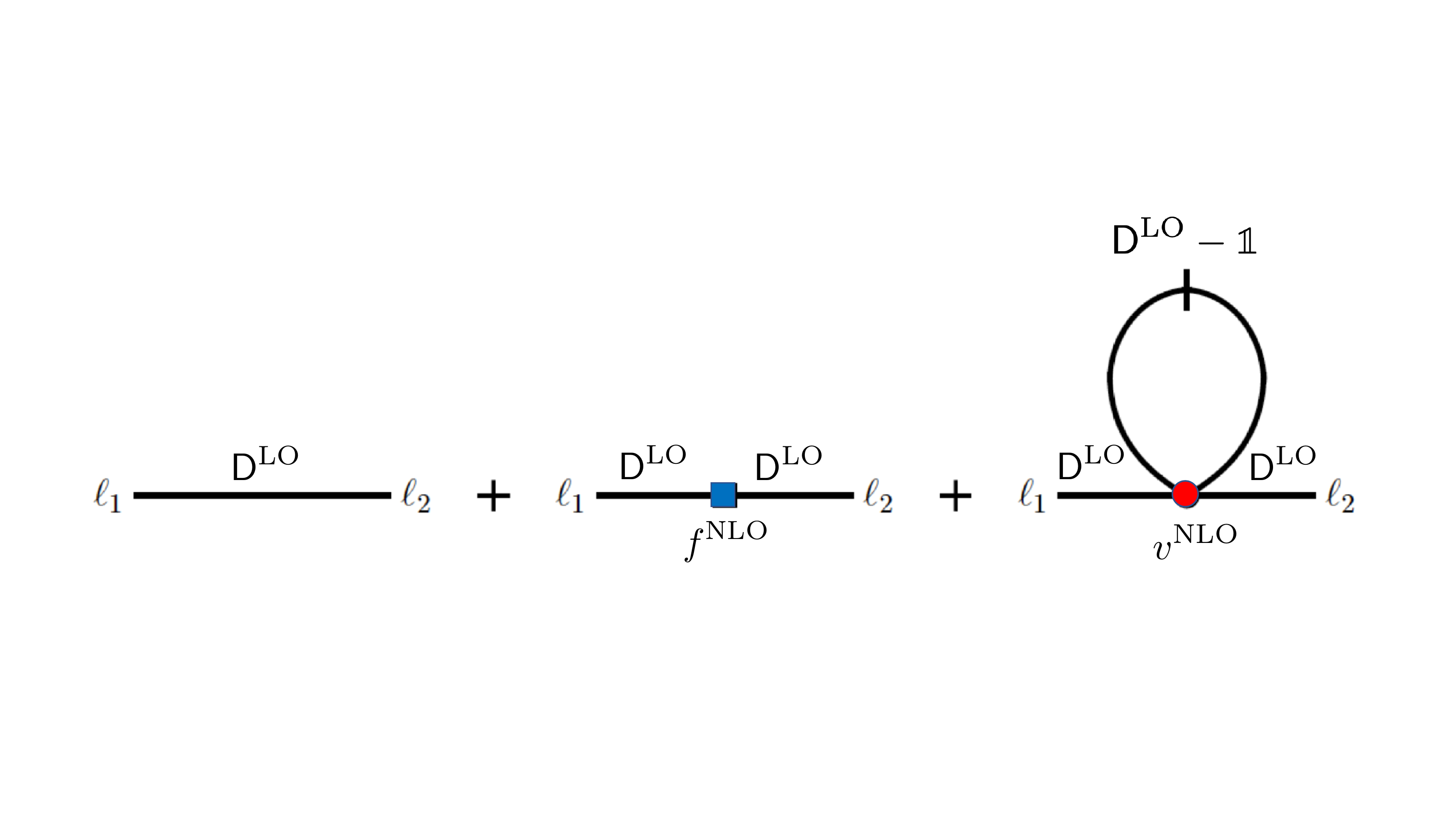}
}
\vspace{-2cm}
\caption{Graphical representation of the right hand side of (\ref{2pointINTfull}). 
The first term corresponds to the LO ``propagator'' $\mathsf{D}^{\mathrm{LO}}$ 
defined in (\ref{2pointINT}). The second term
represents the NLO contribution to the propagator arising from the quadratic ``vertex'' $f^{\mathrm{NLO}}$
defined in (\ref{f}) which we have drawn as a blue square. The last term describes the contribution due to the quartic ``vertex'' $v^{\mathrm{NLO}}$ given in (\ref{Rvertex}) which is represented by a red circle. Note that all lines stand for the propagator $\mathsf{D}^{\mathrm{LO}}$. The only exception is the closed line above $v^{\mathrm{NLO}}$, which represents the difference
$\mathsf{D}^{\mathrm{LO}}-\mathbb{1}$.}
\label{fig:10}
\end{figure}

When we include the
$1/N^2$ corrections, after some algebra we find that (\ref{2pointINT}) becomes
\begin{align}
\label{2pointINTfull}
\big\langle \omega_{2\ell_1+1}\,\omega_{2\ell_2+1}\big \rangle
\,\simeq\, \mathsf{D}^{\mathrm{LO}}
_{\ell_1,\ell_2} 
+ \frac{1}{N^2}\Big( 
\mathsf{D}^{\mathrm{LO}}
_{\ell_1,m}
\,f^{\mathrm{NLO}}_{m,n}\,
\mathsf{D}^{\mathrm{LO}}_{n,\ell_2}
+\frac{1}{2}\,
\mathsf{D}^{\mathrm{LO}}_{\ell_1,k}
\,v^{\mathrm{NLO}}_{k,m,p,q}\,\big(\mathsf{D}^{\mathrm{LO}}_{p,q}-\delta_{p,q}\big)
\,\mathsf{D}^{\mathrm{LO}}
_{m,\ell_2}\Big)\phantom{\bigg|}
\end{align}
where repeated indices are summed over. The right hand side can be graphically represented as
in Fig.~\ref{fig:10} which shows that the sub-leading terms can be understood as corrections to the
``propagator'' $\mathsf{D}^{\mathrm{LO}}$ due to the NLO ``vertices'' $f^{\mathrm{NLO}}$ and
$v^{\mathrm{NLO}}$.

We have checked this result in the perturbative regime for several values of $\ell_1$ and $\ell_2$ by expanding the matrix $\Xx$ for $\lambda\to0$.
On the other hand, knowing that $\mathsf{D}^{\mathrm{LO}}$ is proportional to $1/\lambda$ at strong coupling and observing that the NLO correction in (\ref{2pointINTfull})
is quadratic and cubic in $\mathsf{D}^{\mathrm{LO}}$, we see that the sub-leading correction to the 2-point correlators are of order $1/\lambda^2$ when $\lambda\to\infty$.
Going back to the initial basis of the $\Omega$ operators by inverting the relation (\ref{omega}), we may conclude that also the NLO terms of the 2-point correlators 
$T_{2\ell_1+1,2\ell_2+1}=\big\langle \Omega_{2\ell_1+1}\,\Omega_{2\ell_2+1}\big \rangle$ in the $\mathbf{E}$ theory are proportional to $1/\lambda^2$ at strong coupling.

Like in the 2-point correlator (\ref{2pointINTfull}), also for the higher point correlators of 
$\omega$ operators the NLO contribution is given by Feynman diagrams constructed with the ``propagator''  $\mathsf{D}^{\mathrm{LO}}$ and with at most one quadratic ``vertex'' $f^{\mathrm{NLO}}$ or one quartic  ``vertex'' $v^{\mathrm{NLO}}$, in which 
two lines may also be contracted with $\mathsf{D}^{\mathrm{LO}}-\mathbb{1}$ as in 
Fig.~\ref{fig:10}.

\end{appendix}

%
%
\providecommand{\href}[2]{#2}\begingroup\raggedright\endgroup

\end{document}